\documentclass[journal]{IEEEtran}

\usepackage{xcolor}
\usepackage[normalem]{ulem}
\usepackage{booktabs}
\usepackage{capekCommands}
\usepackage[pdftex]{graphicx}
\usepackage{cuted}
\usepackage{physics}
\usepackage{cite}
\usepackage{pgfplots}
\usepgfplotslibrary{colorbrewer}

\usepackage[]{hyperref}
\hypersetup{
	colorlinks=true, 
	linkcolor=blue!70!black, 
	citecolor=red!70!black, 
	filecolor=blue!70!black, 
	urlcolor=blue!70!black, 
}

\providecommand*{\unit}[1]{\ensuremath{\mathrm{\,#1}}}
\providecommand*{\eu}{\T{e}}

\providecommand{\Umat}{\M{U}} 
\providecommand{\Smat}{\M{U}} 

\providecommand{\aCircum}{r_\srcRegion} 

\graphicspath{{figures/}}

\begin{document}
\title{Unified Theory of Characteristic Modes:\\Part~II -- Tracking, Losses, and FEM Evaluation}
\author{Mats~Gustafsson, \IEEEmembership{Senior Member, IEEE},
Lukas~Jelinek, \\
Kurt~Schab, \IEEEmembership{Member, IEEE}, and
Miloslav~Capek, \IEEEmembership{Senior Member, IEEE}
\thanks{Manuscript received \today; revised \today. This work was supported by the  Swedish Research Council (2017-04656) and Czech Science Foundation under project~\mbox{No.~21-19025M}.}
\thanks{M. Gustafsson is with Lund University, Lund, Sweden (e-mail: mats.gustafsson@eit.lth.se).}
\thanks{L. Jelinek and M. Capek are with the Czech Technical University in Prague, Prague, Czech Republic (e-mails: \{lukas.jelinek; miloslav.capek\}@fel.cvut.cz).}
\thanks{K. Schab is with the Santa Clara University, Santa Clara, USA (e-mail: kschab@scu.edu).}
\thanks{Color versions of one or more of the figures in this paper are
available online at http://ieeexplore.ieee.org.}
\thanks{Digital Object Identifier XXX}
}

\maketitle

\begin{abstract}
This is the second component of a two-part paper dealing with a unification of characteristic mode decomposition. This second part addresses modal tracking and losses and presents several numerical examples for both surface- and volume-based method-of-moment formulations. A new tracking algorithm based on algebraic properties of the transition matrix is developed, achieving excellent precision and requiring a very low number of frequency samples as compared to procedures previously reported in the literature. The transition matrix is further utilized to show that characteristic mode decomposition of lossy objects fails to deliver orthogonal far fields and to demonstrate how characteristic modes can be evaluated using the finite element method.  
\end{abstract}


\IEEEpeerreviewmaketitle

\section{Introduction}
\label{Sec:Introduction}

\IEEEPARstart{P}{art}~I of this two-part paper introduced the theoretical principles of the unified theory of characteristic modes~\cite{Gustafsson+etal_CMT1_2021}, mathematically connecting the impedance-~\cite{HarringtonMautz_TheoryOfCharacteristicModesForConductingBodies} and scattering-based~\cite{Garbacz_TCMdissertation} approaches to characteristic mode decomposition. This second part develops a procedure for modal tracking based on scattering data interpolation, discusses characteristic modes of lossy systems and their properties, and demonstrates the evaluation of characteristic modes using the finite element method (FEM). All parts of the paper are supplemented by examples and benchmarking.  Throughout this paper, the theoretical framework developed in Part~I is utilized~\cite{Gustafsson+etal_CMT1_2021}.

Modal tracking~\cite{CapekHazdraHamouzEichler_AMethodForTrackingCharNumbersAndVectors, RainesRojas_WidebandCharacteristicModeTracking, LudickJakobusVogel_AtrackingAlgorithmForTheEigenvectorsCalculatedWithCM, SafinManteuffel_AdvancedEigenvalueTrackingofCM, Maseketal_ModalTrackingBasedOnGroupTheory,SchabEtAl_EigenvalueCrossingAvoidanceInCM} is a challenging post-processing step required both for visual inspection of the eigentraces~\cite{SchabEtAl_EigenvalueCrossingAvoidanceInCM} (characteristic numbers parameterized by frequency) and interpretation of physical properties of characteristic modes~\cite{HarringtonMautz_ControlOfRadarScatteringByReactiveLoading, Maseketal_ModalTrackingBasedOnGroupTheory}. In this work, decomposition via ordinary eigenvalue problems involving matrices with favorable algebraic properties is used to develop a tracking algorithm which operates natively over far fields and utilizes approximations of poles in the complex frequency plane. As a result, the procedure is accurate, computationally inexpensive, simple, and gives excellent results with regards to precision and the low number of frequency samples needed to reconstruct smooth, broadband eigentraces.

Issues resulting from attempts to define characteristic modes for lossy objects, \eg{},~\cite{HarringtonMautzChang_CharacteristicModesForDielectricAndMagneticBodies} are examined from the perspective of unified scattering- and impedance-based characteristic mode theory. Thanks to the direct link between the impedance and transition matrices, the physical and algebraic properties of the latter are employed to demonstrate that, for lossy scatterers, it is in general not possible to acquire orthogonal far fields and simultaneously diagonalize the matrix used for the decomposition. This is a fundamental observation which applies to all definitions of characteristic modes for lossy objects, \eg{},~\cite{HarringtonMautzChang_CharacteristicModesForDielectricAndMagneticBodies}. 

The scattering-based decomposition utilizing transition matrix is independent of numerical method used~\cite{Gustafsson+etal_CMT1_2021}. This feature is demonstrated with the finite element method (FEM)~\cite{Demesy+etal2018,Fruhnert+etal2017} applied to characteristic mode decomposition of a dielectric cylinder. The scattering-based formalism renders characteristic mode theory a general frequency-domain method applicable to various problems, including non-homogeneous dielectrics, in which methods based on differential equations excel.

The paper is organized as follows.  Modal tracking and various interpolation schemes are shown and discussed in  Sections~\ref{sec:tracking} and~\ref{sec:interp}. Their validity is tested on two numerical examples of a PEC plate and a dielectric cylinder. Properties of characteristic decomposition in lossy systems are discussed Section~\ref{sec:CML}. Section~\ref{S:FEM} serves as a proof of concept of evaluating characteristic modes using the finite element method. The paper is concluded in Section~\ref{sec:concl}.

\section{Mode Tracking and Phase Interpolation}
\label{sec:tracking}

Characteristic mode decomposition is solved independently for each frequency, \ie{}, the characteristic numbers and characteristic modes are initially uncorrelated at discrete frequency points, though it is often desirable to apply modal tracking to order and associate modal quantities such that they may be interpreted as continuous functions of frequency.  Many methods have been proposed for numerical \cite{CapekHazdraHamouzEichler_AMethodForTrackingCharNumbersAndVectors, RainesRojas_WidebandCharacteristicModeTracking, LudickJakobusVogel_AtrackingAlgorithmForTheEigenvectorsCalculatedWithCM, MiersLau_WideBandCMtrackingUtilizingFarFieldPatterns, SafinManteuffel_AdvancedEigenvalueTrackingofCM, Chen2021} or analytical symmetry-based \cite{SchabBernhard_GroupTheoryForCMA,Maseketal_ModalTrackingBasedOnGroupTheory} tracking of characteristic mode quantities, however the small size of the transition matrix~$\M{T}$~\cite{Waterman1965,Gustafsson+etal_CMT1_2021} and the high accuracy of its corresponding characteristic mode decomposition with eigenvectors related to the far field, see Part~I~\cite{Gustafsson+etal_CMT1_2021}, make it exceptionally well suited for this task. 

Consider a case where the transition matrix~$\M{T}(k_q)$ is computed for a set of wavenumbers (frequencies) $k_q$ for~$q=\left\{1,2,\dots,N_{\T{q}}\right\}$. At each wavenumber its eigenvalues~$t_n(k_q)$ and eigenvectors~$\M{f}_n(k_q)$ are calculated via
\begin{equation}
   \label{eq:CM5}
   \M{T}  \M{f}_{n} = t_n \M{f}_{n},
\end{equation}
where in method-of-moments-based formulations the matrix $\M{T}$ is determined from the impedance matrix $\M{Z}$ and projection matrix $\M{U}$ as
\begin{equation}
\label{eq:IZV2}
\M{T} = - \Umat \M{Z}^{-1} \Umat^\T{T},
\end{equation}
see Part~I for details~\cite{Gustafsson+etal_CMT1_2021}. We recall here that eigenvalues~$t_n$ are associated with characteristic numbers~$\lambda_n$ as
\begin{equation}
\label{eq:tValues}
  t_n = -\frac{1}{1 + \T{j}\lambda_n},
\end{equation}
eigenvalue magnitudes~$|t_n|$ are commonly referred to as characteristic modal significance \cite{Austin_1998_TCM_NVIS}, and eigenvectors~$\M{f}_n$ are the coefficients of field expansion into spherical vector waves.

Tracking based on maximizing correlation between each eigenvector $\M{f}_m$ at frequency $k_q$ and eigenvectors $\M{f}_n$ at frequency $k_{q+1}$, \ie{},
\begin{equation}
    \max_{n}\{|\M{f}_m(k_q)^{\herm}\M{f}_n(k_{q+1})|\}
\label{eq:track3}
\end{equation} 
is used here. The eigenvectors $\M{f}_n$ contain the expansion coefficients of the far field $\V{F}_n$ and by means of (see Part~I~\cite{Gustafsson+etal_CMT1_2021})
\begin{equation}
\label{eq:FForth}
\M{f}^\herm_m(k_q) \M{f}_n(k_{q+1})= \frac{1}{Z} \int \limits_{4 \pi} \V{F}_m^*(\hat{\V{r}},k_q) \cdot \V{F}_n(\hat{\V{r}},k_{q+1}) \D{\Omega},
\end{equation}
with free space impedance $Z$, the procedure in \eqref{eq:track3} corresponds to far-field tracking~\cite{MiersLau_WideBandCMtrackingUtilizingFarFieldPatterns,SafinManteuffel_AdvancedEigenvalueTrackingofCM}.

Assuming tracked modes~\eqref{eq:track3} with eigenvalues $t_n=(s_n-1)/2$ expressed as scattering parameters
\begin{equation}
    s_n(k_q)= \exp\left\{\T{j}\phi_n(k_q)\right\}
    \label{eq:sphase}
\end{equation} 
on a unit circle, see Part~I~\cite{Gustafsson+etal_CMT1_2021}, a phase~$\phi_n$ wrapped to the interval~$[-\pi,\pi]$ has discontinuities and is hence not suitable for interpolation. The phase $\phi_n$ is, however, $2\pi$-periodic and unwrapping can be used to remove the jumps and produce a smooth function suitable for interpolation. 

\subsection{Example: A PEC Plate}
\label{subsec:PECplate}

\begin{figure}[]
    \centering
    \includegraphics[width=\columnwidth]{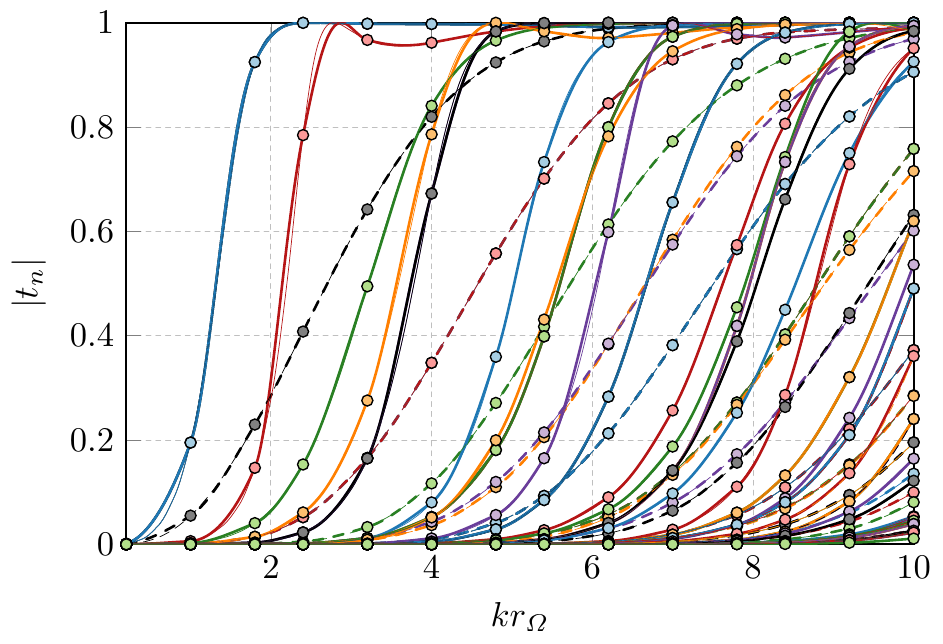}    
    \includegraphics[width=\columnwidth]{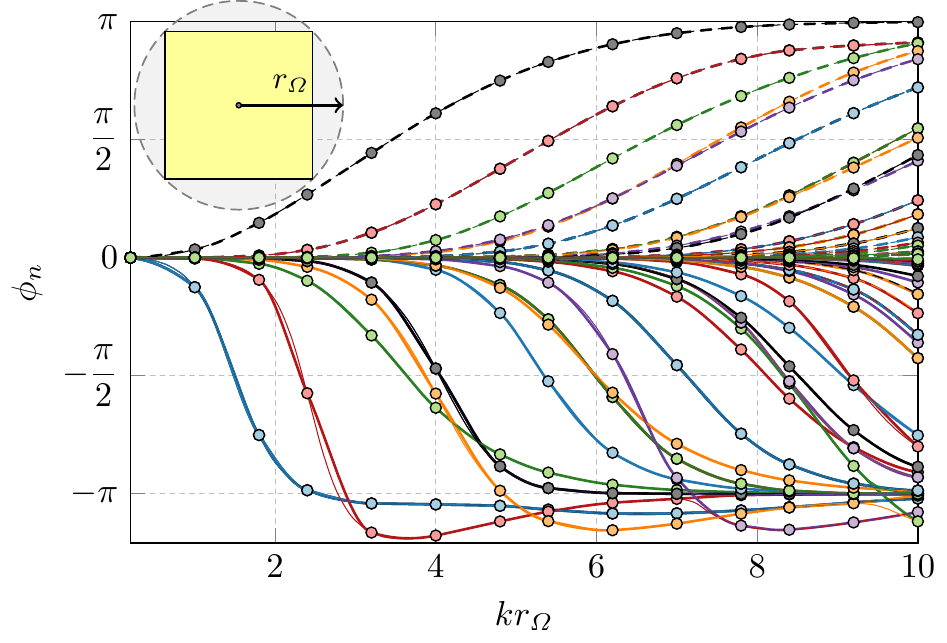}
    \caption{Modal significance $|t_n|$ (top) and phase $\phi_n$ (bottom) for a square PEC plate computed from 14 equidistant sample points (markers) using correlation tracking~\eqref{eq:track3} and higher order interpolation of the unwrapped phase~\eqref{eq:sphase} (thick lines). The data are compared with the corresponding case using 50 sample points (thin lines). The solid lines depict capacitive modes, while the dashed lines depict inductive modes.}
    \label{fig:ResPECplate}
\end{figure}

A square perfectly electric conducting (PEC) plate with side length $\sqrt{2}\aCircum$ is used to illustrate the capabilities of the tracking and higher-order interpolation of the unwrapped phase~$\phi_n$ in~\eqref{eq:sphase}. The impedance matrix~$\M{Z}$ is computed at 14 sample points equidistantly placed in the interval~$k\aCircum\in[0.1,10]$ using the electric field integral equation (EFIE) with 7080 basis functions, see Fig.~\ref{fig:ResPECplate}.

The maximum necessary degree of spherical vector waves
\begin{equation}
    \label{eq:Lmax}
    L_\T{max} = \lceil k\aCircum+\iota\sqrt[3]{k\aCircum}+3 \rceil,
\end{equation}
with~$\iota=7$ used in Part~I is conservative and in many practical applications it is sufficient to compute only a few of the lowest-order characteristic modes, \ie{}, those with highest modal significance~$|t_n|$. The examples presented in this paper were computed with $\iota=7$ and $\iota=2$ and the differences were negligible for modes with modal significance $|t_n| > 0.01$.
Parameters~$k \aCircum=10$ and $\iota=2$ were therefore used in~\eqref{eq:Lmax} giving $L_\T{max}=18$ and $720$~spherical waves. 

The phases~$\phi_n$ of modes tracked via~\eqref{eq:track3} were unwrapped, interpolated using Akima interpolation~\cite{Akima1970}, and are plotted in Fig.~\ref{fig:ResPECplate}. The phase values resulting from~\eqref{eq:sphase} 
are highlighted by markers. For ease of presentation, the unwrapped modal phases~$\{\phi_n\}$ are plotted outside the usual interval~$\left(-\pi,\pi \right)$. It is nevertheless important to realize that the phase $\phi_n$ is ambiguous by addition of multiples of $2\pi$ and that different phase plots might correspond to the same physical scenario unambiguously characterized by the complex number~$s_n$ or~$t_n$. Extreme cases of~$\phi_n = \pm \pi$ are physically the same and correspond to external resonance. 

Evaluation of the results in Fig.~\ref{fig:ResPECplate} from precomputed matrices $\M{Z}$ and $\M{U}$ took $60\unit{s}$ (using MATLAB and i7-9700k CPU @ 3.6GHz with 64\,GB RAM) and was dominated by $59\unit{s}$ used for evaluation of matrix $\M{T}$ via relation~\eqref{eq:IZV2} in the 14 sample points. In contrast, only~$1\unit{s}$ was required for eigenvalue decomposition of the matrix $\M{T}$, tracking, and interpolation. The same time of $60\unit{s}$ would be needed for the evaluation of only 10 dominant characteristic modes using the generalized eigenvalue decomposition
\begin{equation}
    \label{eq:CMI}
    \M{X}_0\M{I}_n = \lambda_n \M{R}_0\M{I}_n,
\end{equation}
with~$\M{R}_0$ and  $\M{X}_0$ being the real and imaginary parts of impedance matrix~\cite{HarringtonMautz_ComputationOfCharacteristicModesForConductingBodies}, in the 14 sample points using $\mathtt{eigs}()$ function in MATLAB. Reducing the discretization to 3120 basis functions reduces the aforementioned evaluation time to $6\unit{s}$ using matrix~$\M{T}$ and to $8\unit{s}$ using the decomposition~\eqref{eq:CMI}. These results indicate that the evaluation of the matrix~$\M{T}$ using~\eqref{eq:IZV2} has a computational cost similar to the cost of evaluating a few dominant characteristic modes using~\eqref{eq:CMI}. The additional computational cost of eigenvalue decomposition~\eqref{eq:CM5}, correlation tracking~\eqref{eq:track3}, and interpolation is negligible, making the scattering-based approach superior in cases when characteristic numbers are to be calculated over broad frequency ranges.

\subsection{Discussion}

Modal tracking using eigenvectors $\M{f}_n$ corresponds to far-field tracking~\cite{MiersLau_WideBandCMtrackingUtilizingFarFieldPatterns,SafinManteuffel_AdvancedEigenvalueTrackingofCM} and benefits from orthogonally of the eigenvectors. This tracking is robust and only requires minor computational overhead from the evaluation of the correlation matrix. Tracking connects modal significances and angles~$\phi_n$ over a frequency band by, \eg{}, drawing straight lines between the sample points. This linear interpolation is improved by higher-order interpolation of the unwrapped phase $\phi_n$. Higher-order interpolation enables the calculation of modes from fewer sampling points and hence faster evaluation of modes over a bandwidth. However, the sampling still needs to be sufficiently dense to resolve resonances and allow for correct tracking. In the next section, an approach for interpolation of the scattering matrix is proposed which resolves these issues.  

\section{Scattering Matrix Interpolation}
\label{sec:interp}
The small size of transition matrix enables direct interpolation of matrix $\M{T}(k)$ instead of interpolation of tracked modes in Section~\ref{sec:tracking}. This is particularly favorable for resonant structures where analytic properties of the scattering matrix
\begin{equation}
    \label{eq:Sdef}
    \M{S} = \M{1}+2\M{T},
\end{equation}
see Part I~\cite{Gustafsson+etal_CMT1_2021}, can be used to obtain low-parameter models of $\M{S}(k)$ and $\M{T}(k)$.  

\subsection{Foster-Based Rational Approximations}

Scattering matrices~\eqref{eq:Sdef}
can be interpolated by fitting rational functions with methods such as Nevanlinna–Pick interpolation~\cite{Delsarte+etal1981} and vector fitting~\cite{1999_Gustavsen_TPD}. For that, the scattering matrix in~\eqref{eq:Sdef} must be made passive~\cite{Zemanian1965} using a time shift~$2\aCircum/c$, where~$c$ denotes the speed of light. This corresponds to multiplication of the scattering matrix~\eqref{eq:Sdef} by a phase term~$\T{exp}\{-2\T{j}k\aCircum\}$, constructing a passive scattering matrix~\cite{Bernland+etal2011b}
\begin{equation}
    \widetilde{\M{S}}=\M{S}\T{e}^{-2\T{j} k\aCircum}.
    \label{eq:Sphaseshift}
\end{equation}
The phase-shifted scattering matrix may be transformed into a lossless transfer matrix~\cite[Chap.~4]{Collin_FoundationsForMicrowaveEngineering} \begin{equation}
	\widetilde{\M{H}} = (\M{1}-\widetilde{\Smat})^{-1}(\M{1}+\widetilde{\Smat}),
\label{eq:track2}
\end{equation}
which is characterized by its poles located at the real axis of the wavenumber~$k$ and its monotonically increasing imaginary part  (Foster's reactance theorem~\cite[Chap.~4]{Collin_FoundationsForMicrowaveEngineering}) for real frequencies between these poles. Quadratic forms $\tilde{h}_y(k)=-\J\M{y}^{\herm}\widetilde{\M{H}}\M{y}$ with arbitrary frequency independent vectors~$\M{y}$ reduce the matrix~$\widetilde{\M{H}}$ to a set of scalar lossless transfer functions $\tilde{h}_y(k)$~\cite{Gustafsson2003}, which can be represented as
\begin{equation}
		\tilde{h}_y(k) = \frac{-b_0}{k} + \sum_{m=1}^M\frac{2b_m k}{\xi_m^2-k^2} + \frac{k}{b_{\infty}},
	\label{eq:Xlossless}
\end{equation}
where $\xi_m\geq 0$, $b_m\geq 0$, and $M$ denotes the number of poles in the frequency interval $k \in (0,\infty)$. Sample points with decreasing imaginary part~$\T{Im}\{\tilde{h}_y(k) \}$ identify intervals where it is necessary to place poles $\xi_m$. There can also be poles outside the range of sample points and in the numerical examples in Section~\ref{S:ResCyl} a single pole in the interval~$(k_{N_{\T{q}}},\infty)$ was used. Locations and residues of the poles are determined by minimizing the least squared error between the rational model~\eqref{eq:Xlossless} and values of the transfer function~$\tilde{h}_y(k)$ at the sample frequencies $\{k_q\}$. It is sufficient to use vectors~$\M{y}$ each having a single non-zero element (pointing to diagonal elements of matrix~$\widetilde{\M{H}}$) together with all vectors~$\M{y}$ having two complex-valued non-zero elements, \eg{}, $1$ and $\J$~\cite{Gustafsson2003} to reconstruct the matrix~$\widetilde{\M{H}}$. The scalar rational functions are used to synthesize a rational matrix function approximating the transfer matrix~$\widetilde{\M{H}}$, which is used to approximate scattering and transition matrices $\M{S}$ and $\M{T}$ between the original sample points from 
\begin{equation}
    \M{S}(k)
    =\eu^{2\T{j}k\aCircum}
    \big(\widetilde{\M{H}}(k)-\M{1}\big)\big(\widetilde{\M{H}}(k)+\M{1}\big)^{-1}.
\label{eq:SfromH}
\end{equation}
This matrix function can hence be used to approximate eigenvalues~$s_n$ and $t_n$  as well as eigenvectors~$\M{f}_n$ in the frequency interval given by the sample points. However, constructing the eigencurrents $\M{I}_n$~using Part~I~\cite[Eq.~(35)]{Gustafsson+etal_CMT1_2021}, requires additional interpolation or evaluation of the impedance matrix $\M{Z}$ at the interpolation frequency points.
The rational fit works best for scatterers with narrowband resonances as illustrated in Section~\ref{S:ResCyl}. For wideband resonances and modes with low modal significance~$|t_n|\approx 0$ a different technique described in next section is more suitable. 

\subsection{Hybridization with Linear Phase Interpolation}

Foster's reactance theorem~\cite[Chap.~4]{Collin_FoundationsForMicrowaveEngineering} can alternatively be formulated as a monotonically decreasing phase of the scattering matrix~$\widetilde{\M{S}}$, with the representation
\begin{equation}
    \widetilde{\M{S}}(k) = \exp(\J\M{\Phi}(k)),
\end{equation}
where $\M{\Phi}(k)$ is a monotonically decreasing Hermitian matrix function\footnote{Defined as having $\M{I}^{\herm}\M{\Phi}(k)\M{I}$ monotonically decreasing for all $\M{I}$.} and $\exp(\cdot)$ denotes the matrix exponent. These phase matrices are related to an unwrapped matrix logarithm of the phase-shifted scattering matrix~$\widetilde{\M{S}}$, similar to standard phase unwrapping for scalar phase functions. Linear interpolation of the phase matrices is also similar to an interpolation between two scattering matrices~$\Smat(k_q)$ and~$\Smat(k_{q+1})$ given by
\begin{equation}
	\Smat(k_q+\nu(k_{q+1}-k_q))  =\Smat(k_q)\big(\Smat(k_q)^{-1}\Smat(k_{q+1})\big)^{\nu},
\label{eq:track1}
\end{equation}
with $0\leq \nu\leq 1$, see Appendix~\ref{S:IntUnitaryMat}. This linear interpolation of the phase works well for non-resonant cases, where the phase changes slowly. 

The two approximations~\eqref{eq:Xlossless}  and~\eqref{eq:track1} are hence complementary, which calls for their combination. Construct first a rational approximation $\M{S}_1$ using~\eqref{eq:SfromH} of the scattering matrix based on dominant elements of the matrix $\M{T}$, \eg{}, large diagonal elements, to capture resonances. 
Removing these resonances from the scattering matrix by dividing $\M{S}$ with $\M{S}_1$ constructs a unitary matrix  
\begin{equation}
    \M{S}_2=\M{S}_1^{-1}\M{S}
    \label{eq:Smatdiv}
\end{equation}
with slower phase variation, which can be interpolated using~\eqref{eq:track1}. The interpolated scattering matrix is finally synthesized as $\M{S}_1\M{S}_2$.    

\subsection{Foster-based phase unwrapping of tracked modes}
Standard unwrapping by adding multiples of $2\pi$ to minimize jumps works well for sampling which is sufficiently dense to resolve all resonances as in Section~\ref{subsec:PECplate}, but does not work for coarse sampling. Shifting the phase according to~\eqref{eq:Sphaseshift} transforms eigenvalues $s_n(k)$ to a lossless scattering parameter $\tilde{s}_n(k) = s_n(k)\exp\left\{-2\T{j}k\aCircum\right\}$ which has a monotonically decreasing phase according to Foster's reactance theorem. This approximation can be used in the unwrapping algorithm by subtracting $2\pi$ from the phase at samples with increasing $\arg(\tilde{s}_n(k_q))$, \ie{}, the unwrapped phase is constructed to satisfy
\begin{equation}
    \phi_n(k_p)-2k_p\aCircum > \phi_n(k_q)-2k_q\aCircum
    \quad\text{for }
    k_q>k_p.
    \label{eq:unwrapping}
\end{equation}
Linear or higher-order interpolation is used to estimate phase between the sample points. This scalar interpolation can hence be used to find scattering resonances~$\phi_n=-\pi+2m\pi$ ($t_n= -1$ or $\lambda_n=0$) for cases when the resonance frequency is not sampled, similarly to the rational fitting discussed above.



\subsection{Example: Dielectric Cylinder}\label{S:ResCyl}
\begin{figure}[]
    \centering
    \includegraphics[width=\columnwidth]{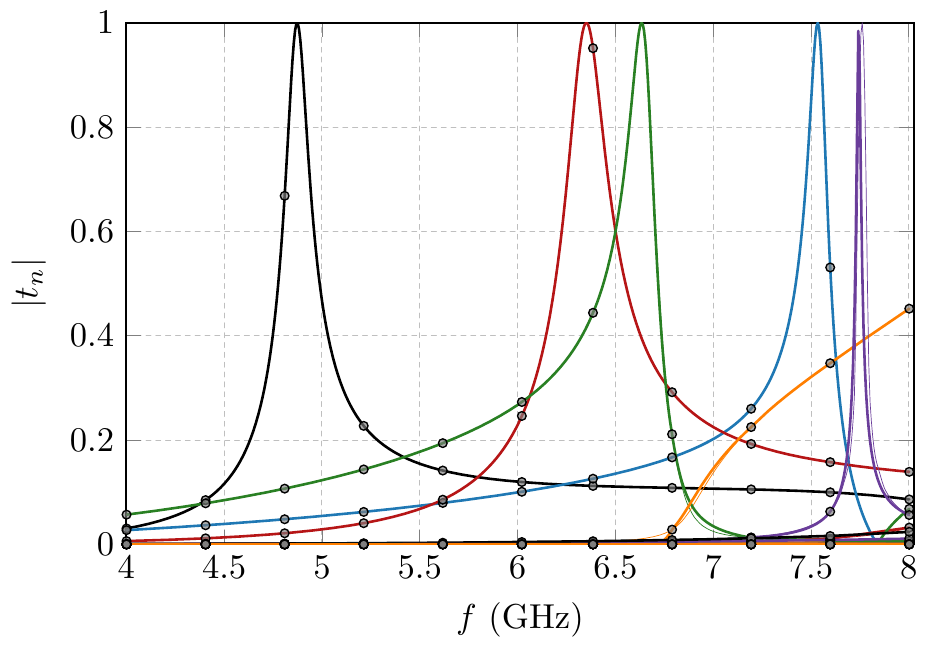}
    \includegraphics[width=\columnwidth]{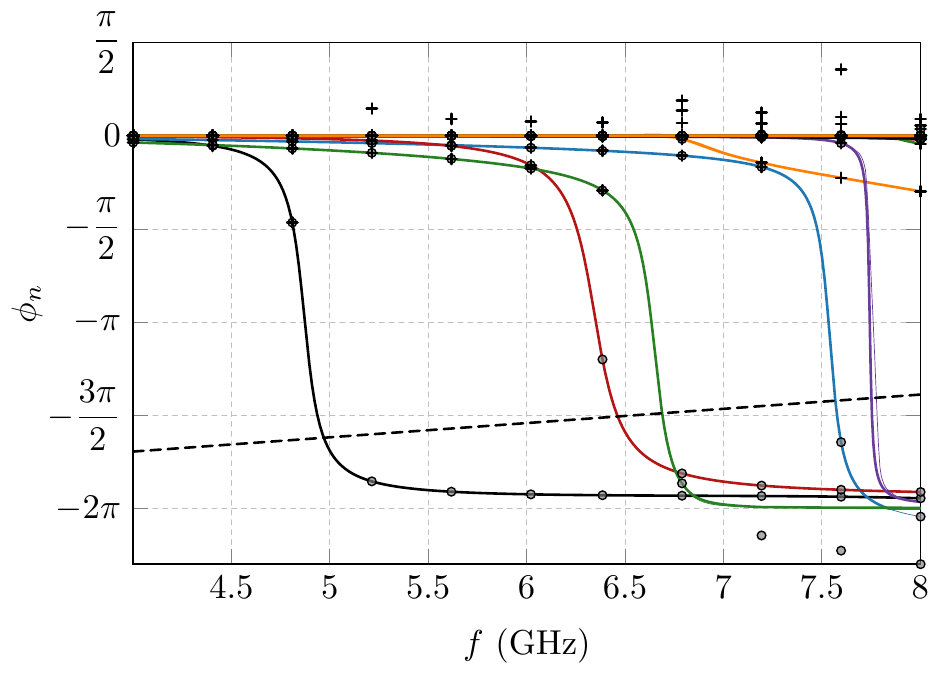}    

    \caption{Modal significance $|t_n|$ (top) and phase $\phi_n$ (bottom) for a dielectric cylinder with height $4.6\unit{mm}$, radius $5.25\unit{mm}$, and relative permittivity $\varepsilon_{\T{r}}=38$ calculated using~\eqref{eq:IZV2} with a PMCHWT formulation. Thick lines show interpolation based on rational functions~\eqref{eq:Xlossless} together with~\eqref{eq:track1} using 11 sample points which are indicated with markers. Thin lines are based on 100 sample points and phase unwrapping~\eqref{eq:unwrapping}. Intersection between the phase $\phi_n$ and the dashed line approximates the poles $\xi_m$ of the rational function~\eqref{eq:Xlossless}.}
    \label{fig:Cyl38absT}
\end{figure}

\begin{figure}[]
    \centering
    

    \includegraphics[width=0.19\columnwidth]{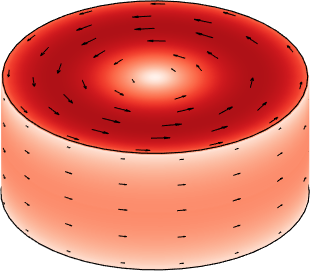}
    \includegraphics[width=0.19\columnwidth]{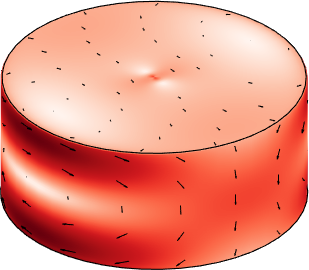}
    \includegraphics[width=0.19\columnwidth]{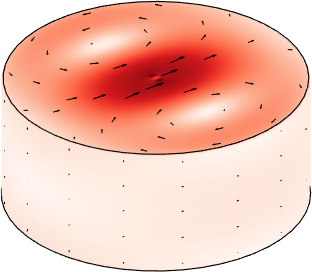}
    \includegraphics[width=0.19\columnwidth]{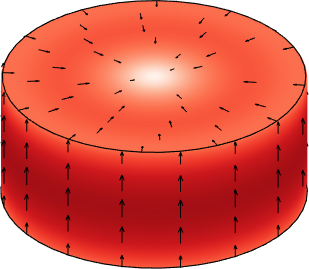}
    \includegraphics[width=0.19\columnwidth]{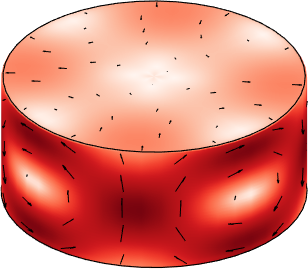}%
    \vspace{1mm}
    
    \includegraphics[width=0.19\columnwidth]{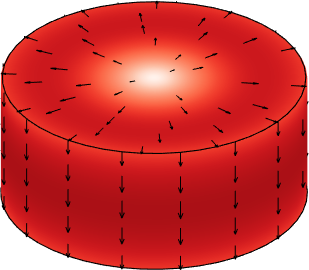}
    \includegraphics[width=0.19\columnwidth]{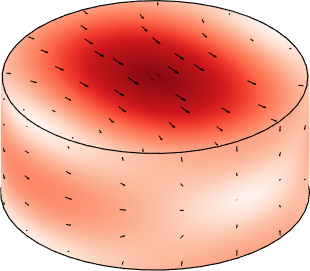}
    \includegraphics[width=0.19\columnwidth]{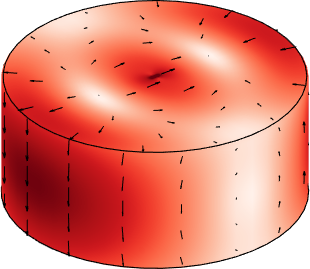}
    \includegraphics[width=0.19\columnwidth]{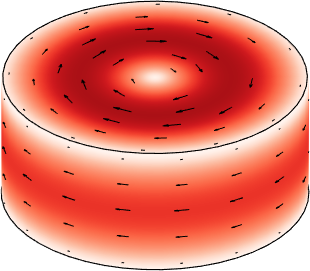}
    \includegraphics[width=0.19\columnwidth]{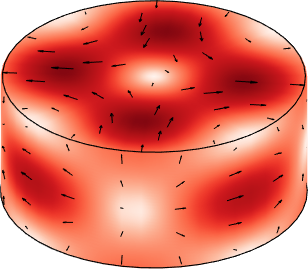}%
    
    {\footnotesize \hspace{5mm} TE$_{01}$\hfill HEM$_{11}$\hfill HEM$_{12}$\hfill TM$_{01}$\hfill HEM$_{21}$\hspace{3mm}} 
    \caption{Characteristic current densities~$\V{J}_n$ (top) and~$\V{M}_n$ (bottom) for the resonance frequencies of the dielectric cylinder in Fig.~\ref{fig:Cyl38absT} computed from the eigenvectors~$\M{f}_n$ using~\eqref{eq:CMTJ}. From left to right, the depicted currents correspond to modes: TE$_{01}$ at $4.87\unit{GHz}$, HEM$_{11}$ at $6.35\unit{GHz}$, HEM$_{12}$ at $6.63\unit{GHz}$, TM$_{01}$ at $7.53\unit{GHz}$, and HEM$_{21}$ at $7.76\unit{GHz}$.}
    \label{fig:Cyl38JM}
\end{figure}

The benchmark example from~\cite{2016_Hu_TAP,Ylae-Oijala2018,Huang+etal2019} consisting of a cylindrical dielectric resonator with height $4.6\unit{mm}$, radius $5.25\unit{mm}$, and relative permittivity $\varepsilon_{\mathrm{r}}=38$ is used to illustrate the potential of the proposed methodology on a challenging and well-studied case. The PMCHWT~\cite{1973_Poggio_ComputerTechniquesForElectromagnetics,1977_Wu_RS,ChangHarrington_AsurfaceFormulationForCharacteristicModesOfMaterialBodies} formulation is used to obtain the impedance matrix. The modal significance $|t_n|$ and phase $\phi_n$ evaluated from~\eqref{eq:CM5} and~$s_n = 1 + 2 t_n$ are depicted in Fig.~\ref{fig:Cyl38absT} showing five distinct peaks corresponding to TE$_{01}$, HEM$_{11}$, HEM$_{12}$, TM$_{01}$, and HEM$_{21}$ modes, in agreement with the results in~\cite{Ylae-Oijala2018,Huang+etal2019}. The  corresponding surface currents calculated from
\begin{equation}
    \M{I}_n
    =t_n^{-1}\M{Z}^{-1}\M{U}^{\T{T}}\M{f}_n,
    \label{eq:CMTJ}
\end{equation}
see Part I~\cite{Gustafsson+etal_CMT1_2021},
at the resonance frequencies are depicted in~Fig.~\ref{fig:Cyl38JM}.

The results in Fig.~\ref{fig:Cyl38absT} also illustrate interpolation using rational functions~\eqref{eq:Xlossless} together with~\eqref{eq:track1}. The rational fit uses 11 equidistant frequency points, shown with black markers, to compute the poles and residues in Table~\ref{tab:ratfit} of the dominant diagonal elements $T_{nn}$ corresponding to the peaks in Fig.~\ref{fig:Cyl38absT}. This diagonal matrix is used to construct a scattering matrix $\M{S}_1$ containing dominant contributions of the resonances. These resonances are removed from $\M{S}$ using~\eqref{eq:Smatdiv} to synthesize a scattering matrix $\M{S}_2$ with slower frequency variation which is interpolated using~\eqref{eq:track1}. The resulting curves for modal significance~$|t_n|$ and angle~$\phi_n$ tracked using~\eqref{eq:track3} are shown in Fig.~\ref{fig:Cyl38absT} by solid curves with colors distinguishing different modes. Intersections between the dashed line, illustrating the resonance condition~$\phi_n + 2 k \aCircum = -\pi$ for the phase shifted scattering matrix~\eqref{eq:Sphaseshift}, and curves are related to the resonance frequencies in the rational model~\eqref{eq:Xlossless} as given in Table.~\ref{tab:ratfit}.  
The interpolation results are compared with results based on 100 sample points tracked using~\eqref{eq:track3} and plotted with thin solid lines. This dense sampling was used to resolve narrowband resonances. The thin lines are mostly indistinguishable from the thicker lines, \ie{}, the curves coincide, which validates the rational approximation~\eqref{eq:track2} combined with~\eqref{eq:track1} using~\eqref{eq:Smatdiv}.    
The rational fit~\eqref{eq:Xlossless} finds all peaks, although they are not sampled in the original computation (11 sample points) and only a slight shift (compared with the thin lines) is seen for the $\T{HEM}_{21}$ mode. The interpolation~\eqref{eq:Smatdiv} based on this small number of sampling points causes some errors, \eg{}, the minor disturbances around $6.7\unit{GHz}$. 

Interpolation based on phase unwrapping~\eqref{eq:unwrapping} of tracked modes~\eqref{eq:track3} is also illustrated in the bottom panel of  Fig.~\ref{fig:Cyl38absT}.  
Sampling at only $11$~points misses most peaks and the resulting phase~$\phi_n$ is close to zero for most of the sampled points as shown by the $+$ markers in Fig.~\ref{fig:Cyl38absT}. Modal tracking~\eqref{eq:track3} based on these $11$~sample points and ordinary phase unwrapping does not affect the phase. Nevertheless, making the eigenvalues~$s_n$ passive (assuming non-degenerate eigenvalues) by the phase shift $\T{exp} (-2\T{j} k\aCircum)$, \cf{}~\eqref{eq:Sphaseshift}, and invoking Foster's theorem in the phase unwrapping~\eqref{eq:unwrapping} decreases many sample points $\phi_n(k_q)$ by a factor $-2\pi$ as shown by the corresponding circular markers. This unwrapped phase can be interpolated with standard low or high order algorithms and produces curves crossing the line $\phi_n = -\pi$ at scattering (external) resonances with $t_n=-1$. One notable error in unwrapping of the tracked modes is observed for the HEM$_{12}$ mode, where the tracking (based on 11 samples) misses the onset of a new mode around $6.7\unit{GHz}$ and predicts a phase following the circular markers below $\phi_n = -2\pi$.   

\begin{table}[]
\caption{Coefficients in unit $\unit{m^{-1}}$ of the rational fit~\eqref{eq:Xlossless} of the diagonal elements $T_{nn}$ used together with $b_0=1/\aCircum\approx 174\unit{m^{-1}}$ to calculate the curves in Fig.~\ref{fig:Cyl38absT}.}
\centering
\label{tab:ratfit}
\begin{tabular}{ccccccc}
 & TE$_{01}$ & HEM$_{11}$ & HEM$_{12}$ & TM$_{01}$ & HEM$_{21}$\\ \toprule
$\xi_1$ & 104 & 135 & 140 & 159 & 162  \\
$b_1$ & 4.59 & 4.85 & 1.78 & 1.29 & 0.363  \\
$\xi_2$ & 199 &  & 247 &  &  \\
$b_2$ & 5.44 &  & 51.2 &  &   \\
$b_\infty$ & 1242 & 667 & 380 & 299 & 518 \\ \bottomrule
\end{tabular}
\end{table}

Using $\iota=2$ in~\eqref{eq:Lmax} corresponds to the degree of the spherical waves $L_\T{max}=12$, \ie{}, 336~spherical waves in total. In this configuration, the computational cost is dominated by the calculation of $\M{Z}^{-1}\M{U}^\T{T}$  in~\eqref{eq:IZV2}. The computation of all characteristic numbers, characteristic modes, and their tracking is computationally much less expensive. 

\subsection{Discussion}
Characteristic mode analysis is often performed for sub-wavelengths objects, where interaction with low order spherical waves (dipoles, quadrupoles,\dots) dominates. This results in small scattering and transition matrices with a few dominant elements, which combined with their analytical properties enable efficient interpolation of the matrices over a frequency range. The properties are further improved by adding a phase shift~\eqref{eq:Sphaseshift} which makes them passive~\cite{Zemanian1965} and after transforming them to transfer functions~\eqref{eq:track2} construct matrices with properties similar to classical lossless circuits, where Foster’s reactance theorem shows that the imaginary part is monotonically increasing between simple poles at the frequency axis. Using low parameter models consisting of the locations and residues of the poles allows for the interpolation of narrow resonances. This rational fit~\eqref{eq:Xlossless} is combined with a linear interpolation of the scattering matrix to capture non-resonant components, as seen in Fig.~\ref{fig:Cyl38absT}.      

Analyticity is in general lost for degenerate eigenvalues~\cite{Kato1980}, meaning that Foster’s reactance theorem and rational interpolation must be used with care for tracked modes, \eg{} by adding samples around frequencies, where the phase is unwrapped~\eqref{eq:unwrapping}.

\section{Characteristic Modes of Lossy Scatterers}
\label{sec:CML}

A natural way of extending the theory developed in Part~I of this paper~\cite{Gustafsson+etal_CMT1_2021} is to take into account lossy obstacles.  In this section, we describe the loss of characteristic far-field orthogonality from the perspective of the properties of the transition matrix~$\M{T}$ and impedance matrix~$\M{Z}$ for lossy systems.

\subsection{Normality of the Transition Matrix}

The far-field orthogonality in characteristic modes for lossless systems
\begin{equation}
\M{f}_m^\herm \M{f}_n = \delta_{mn}
\label{eq:fforto}
\end{equation}
followed from the spectral theorem which states that a matrix is unitary diagonalizable if and only if the matrix is normal~\cite{HornJohnson_MatrixAnalysis}.  Justification of the normality of the matrix $\M{T}$ for lossless systems is given in Appendix A of part~I~\cite{Gustafsson+etal_CMT1_2021}.  
In lossy systems, the normality of matrix $\M{T}$ and unitarity of matrix~$\M{S}$ is generally lost which can be linked to the loss of time-reversal symmetry~\cite[Chap.~6]{Jackson_ClassicalElectrodynamics}. Assuming reciprocity $\M{T}^{\trans}=\M{T}$ and using the fact that $\M{T}$ is normal if and only if~\cite{Zhang2011} 
\begin{equation}
     \norm{\M{T}\M{a}}
     =\norm{\M{T}^{\ast}\M{a}}
 \label{eq:NormTR}
 \end{equation}
for all $\M{a}$, with $\ast$ denoting complex conjugate, it is observed that the lack of time-reversal symmetry, $\M{T} \neq \M{T}^*$, leads to the loss of normality unless special circumstances, such as a diagonal matrix~$\M{T}$, occur.

Normality of the transition matrix can further be linked to specific properties of the underlying impedance matrix. Consider, for example, an impedance matrix of the form of $\M{Z}=\M{R}_{\rho}+\M{R}_0+\J\M{X}$ with $\T{Re}\{\M{Z}\}=\M{R}_{\rho}+\M{R}_0$ and $\M{R}_\rho$ representing losses.  The relation~\eqref{eq:IZV2} can be used to rewrite $\M{T}^{\herm}\M{T}=\M{T}\M{T}^{\herm}$ as
\begin{equation}
  \M{U}(\M{Z}^{-\herm}\M{R}_0\M{Z}^{-1})\M{U}^{\trans}
  =\M{U}(\M{Z}^{-1}\M{R}_0\M{Z}^{-\herm})\M{U}^{\trans}
\label{eq:TnormZ},
\end{equation}
or equivalently
\begin{equation}
	\M{U}(\M{Z}^{-\herm}\M{R}_\rho\M{Z}^{-1})\M{U}^{\trans}=\M{U}(\M{Z}^{-1}\M{R}_\rho\M{Z}^{-\herm})\M{U}^{\trans}.
\label{eq:simDiag}
\end{equation}
Assuming further that the matrix~$\M{R}_\rho$ is invertible, the inner part of relation~\eqref{eq:simDiag} simplifies to 
\begin{equation}
    \M{Z}\M{R}_\rho^{-1}\M{Z}^{\herm} = \M{Z}^{\herm}\M{R}_\rho^{-1}\M{Z}
\end{equation}
which, with~$\M{X} = \T{Im} \left\{ \M{Z} \right\}$, reduces to
\begin{equation}
    \M{R}_{0}\M{R}_\rho^{-1}\M{X}
    =\M{X}\M{R}_\rho^{-1}\M{R}_{0}.
    \label{eq:EFIEsimdiag}
\end{equation}
This resembles a commutator relation between $\M{R}_0$ and $\M{X}$ weighted by $\M{R}_{\rho}^{-1}$. The matrices in~\eqref{eq:EFIEsimdiag} can be permuted as long as they are invertible.  

Because the above relation represents a sufficient condition for normality of the transition matrix,
we conclude that it is in general not possible to diagonalize the matrix~$\M{T}$ of a lossy scatterer with eigenmodes having orthogonal far fields.
Rather, as shown in \cite{HarringtonMautzChang_CharacteristicModesForDielectricAndMagneticBodies}, eigenmodes diagonalizing the matrix~$\M{T}$ have (assuming reciprocity) a diagonal matrix of reaction products
\begin{equation}
\M{f}_m^\trans \M{f}_n = c_n\delta_{mn},
\label{eq:ff_reaction}
\end{equation}
where $^\trans$ is the matrix transpose and~$|c_n|\leq 1$ assuming that modes are normalized as~$\M{f}_n^\herm \M{f}_n = 1$. This nevertheless does not guarantee far-field orthogonality as defined in \eqref{eq:FForth} and, in general for lossy systems we have the eigenvectors of the matrix~$\M{T}$ giving
\begin{equation}
\M{f}_m^\herm \M{f}_n = \rho_{mn},
\label{eq:rho}
\end{equation}
with~$\rho_{mn}$ being a matrix of correlation coefficients with non-zero off-diagonal elements.  Note that the magnitude squared of these coefficients correspond to the envelope correlation coefficients used to describe multiple-input multiple-output (MIMO) antenna systems~\cite{1987_Vaughan_VT,thaysen2006envelope}. 

Losses also change the eigenvalues $t_n$ and $s_n$ from being restricted to the rim of the circles in the complex plane (see Part~I~\cite{Gustafsson+etal_CMT1_2021}, Fig.~2) to also occupying their inner regions. Modes satisfying $|s_n|<1$ are interpreted as lossy. The corresponding characteristic values~$\lambda_n$ are then transformed from being real-valued in the lossless case to complex-valued~\cite{HarringtonMautzChang_CharacteristicModesForDielectricAndMagneticBodies} in the lossy case, with modal dissipation factor~\cite{YlaOijala_GeneralizedTCM2019} being given as~$\delta_n = - \T{Im} \left\{\lambda_n \right\}$. These statements follow directly from the realization that cycle mean modal lost power is given by
\begin{equation}
    P_{\T{lost},n} = \T{Re} \left\{ P_{\T{c},n} \right\} - P_{\T{rad},n}
    \label{eq:Plost}
\end{equation}
and therefore
\begin{equation}
    \delta_n = \dfrac{P_{\T{lost},n}}{P_{\T{rad},n}} = - \T{Re} \left\{ 1 + \dfrac{1}{t_n} \right\},
    \label{eq:Plost1}
\end{equation}
see Appendix~A and~formula~(18) in Part~I~\cite{Gustafsson+etal_CMT1_2021}.

\subsection{Interpretation of Simultaneous Diagonalization}

The condition in~\eqref{eq:EFIEsimdiag} is equivalent to the requirement of simultaneous diagonalization (by congruence) of three symmetric matrices $\M{X},\M{R}_0$, and $\M{R}_{\rho}$, see~\cite{Novikov2014} and Appendix~\ref{S:DiagTreeMat}, which is not possible in most cases.
However additional perspectives can be obtained by studying hypothetical cases where simultaneous diagonalization of the three matrices $\M{X}$, $\M{R}_0$, and $\M{R}_\rho$ is possible.  If such diagonalization were possible, then, the following orthogonality relations hold
\begin{align}
\Ivec_m^\herm \M{R}_0 \Ivec_n &= \delta_{mn}, \label{eq:ortho3A} \\
\Ivec_m^\herm \M{R}_\rho \Ivec_n &= d_m \delta_{mn}, \label{eq:ortho3B} \\
\Ivec_m^\herm \M{X} \Ivec_n &= l_m \delta_{mn}. \label{eq:ortho3C}
\end{align}
For homogeneous dielectric bodies, the ohmic loss operator in~\eqref{eq:ortho3B} is generated by the Gramian matrix scaled by a resistivity~$\rho$, \ie{}, $\M{R}_\rho = \RE\left\{\rho\right\} \V{\Psi}$~\cite{GustafssonCapekSchab_TradeOffBetweenAntennaEfficiencyAndQfactor}. Combining orthogonality relations~\eqref{eq:ortho3A} and \eqref{eq:ortho3B} yields
\begin{equation}
\Ivec_m^\herm \left( \M{R}_0 + \V{\Psi} \right) \Ivec_n = \left( d_m / \RE\left\{\rho\right\} + 1 \right) \delta_{mn}.
\label{eq:orthoRL}
\end{equation}
It is obvious that if characteristic modes for lossy objects would exist, their modal currents would not depend on the complex resistivity~$\rho$, \ie{}, they would have the same shape as for an equivalent lossless system, \eg{}, PEC obstacles. Interestingly, this hypothesis can be confirmed for rare cases where the simultaneous diagonalization of~\eqref{eq:ortho3A}--\eqref{eq:ortho3C} is possible, \ie{}, for fully separable bodies~\cite{MorseFeshBach_MethodsOfTheoreticalPhysics} like a spherical shell, for which the characteristic currents are vector spherical harmonics~\cite{Garbacz_TCMdissertation, CapekEtAl_ValidatingCMsolvers}, and their shape is constant both in frequency and for arbitrary resistivity. Only the eigenvalues are properly rescaled.
Other notable exceptions exist in cases involving obstacles with cylindrical and planar symmetries, where~$\M{T}$ is diagonal (and normal) by construction~\cite{MorseFeshBach_MethodsOfTheoreticalPhysics}, and in cases with dominant modes belonging to different irreducible representations of a symmetry group~\cite{McWeeny_GroupTheory, Maseketal_ModalTrackingBasedOnGroupTheory} such as modes with even and odd (reflection) symmetry, see Section~\ref{sec:resDipole}.

\subsection{Example: Lossy Dipoles}
\label{sec:resDipole}
\begin{figure}[]
    \centering
    \includegraphics[width=\columnwidth]{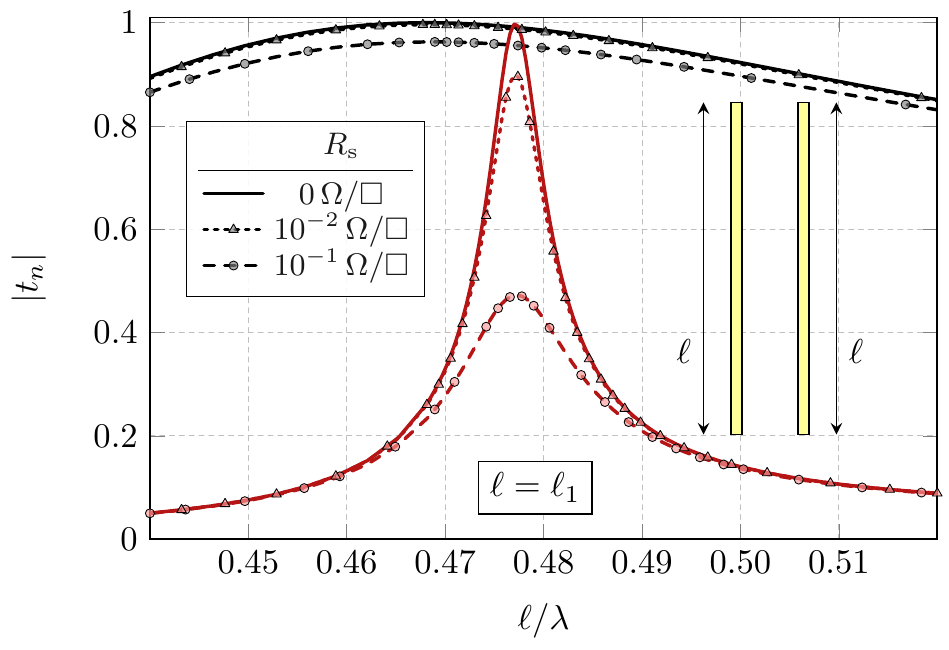}
    \includegraphics[width=\columnwidth]{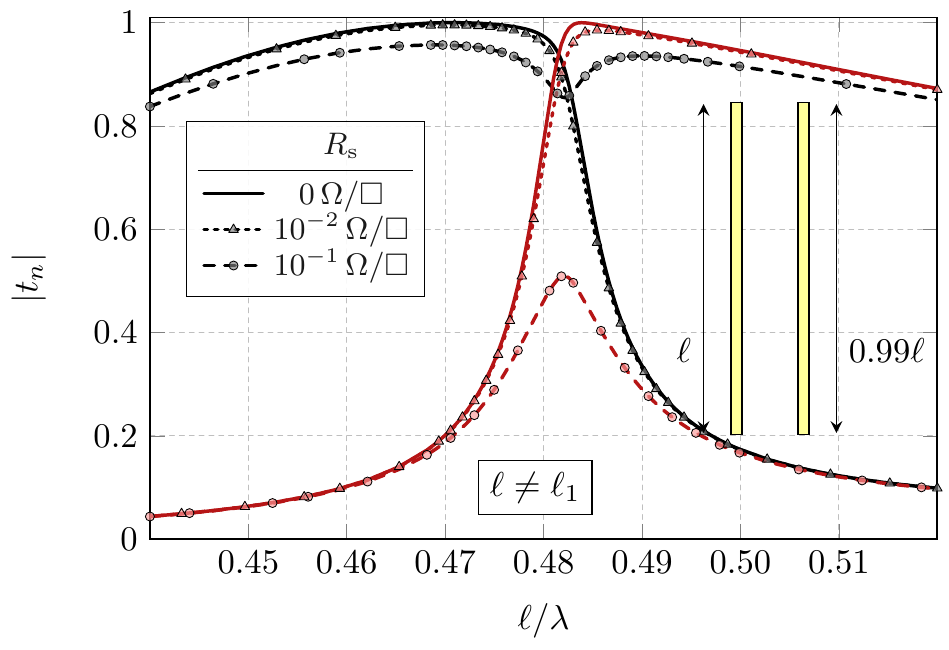}
    \caption{Modal significance for two dipoles with lengths $\ell$ and $\ell_1$, width $\ell/100$, separated by a distance $\ell/5$ (outside to outside) and modeled with surface resistivity $R_{\T{s}}\in\{0,10^{-2},10^{-1}\}\unit{\Omega/\square}$ as indicated by the legend. Two dominant modes (even and odd modes) are shown by black and red curves. Top figure corresponds to identical dipoles $\ell=\ell_1$ while the bottom figure corresponds to lengths $\ell_1=0.99\ell$.}
    \label{fig:DipolesAbsT_final}
\end{figure}

Two thin and closely spaced dipoles are used to illustrate tracking of characteristic modes for structures with resonances and symmetries, as well as to show effects of losses. Consider two strip dipoles with lengths $\ell$ and $\ell_1$, strip widths $0.01\ell$, separated by a distance $0.2\ell$, and modeled by a surface resistivity $R_{\T{s}}\in\{0,10^{-2},10^{-1}\}\unit{\Omega/\square}$ as depicted in the inset of Fig.~\ref{fig:DipolesAbsT_final}.

\begin{figure}[]
    \centering
    \includegraphics[width=\columnwidth]{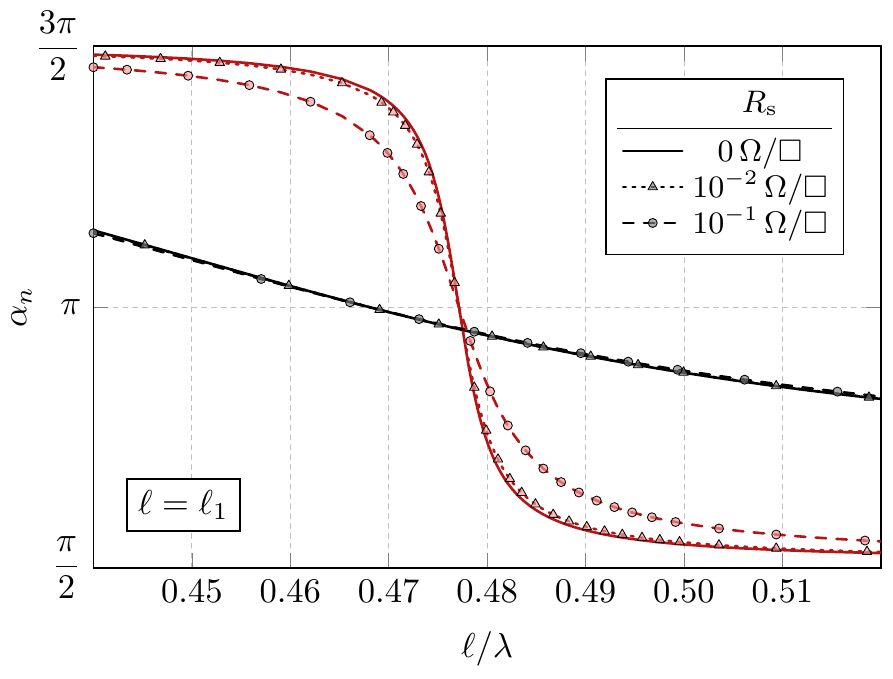}    
    \includegraphics[width=\columnwidth]{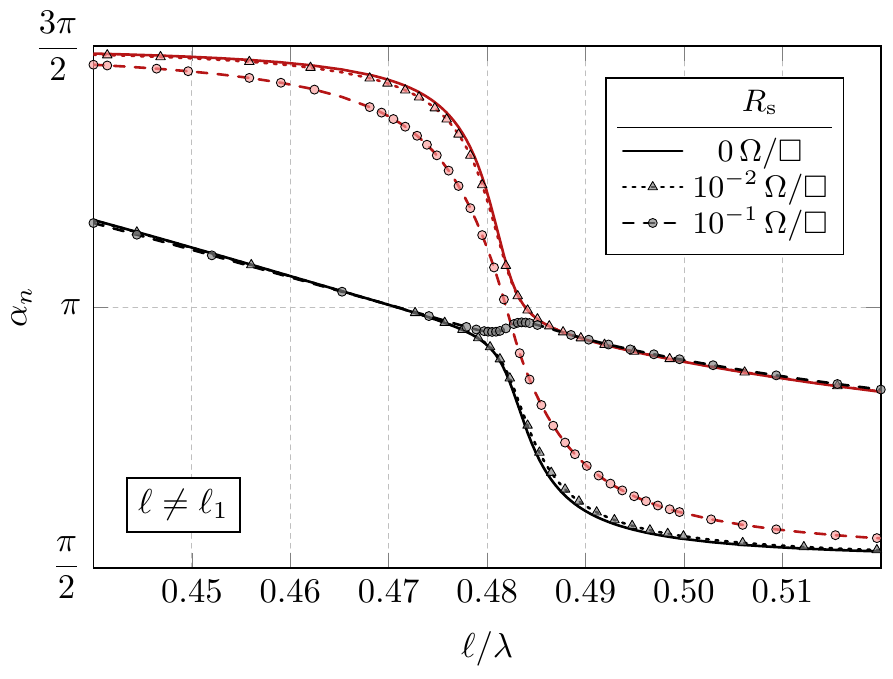}
    \caption{Phase $\alpha_n$ for a setup identical with Fig.~\ref{fig:DipolesAbsT_final}}
    \label{fig:DipolesArgS}
\end{figure}

Lossless structures ($R_{\T{s}}=0$, solid curves in Figs.~\ref{fig:DipolesAbsT_final} and~\ref{fig:DipolesArgS}) are used first to examine the ability of correlation tracking~\eqref{eq:track3} to properly identify eigenvalue crossings and crossing avoidances. Magnitudes~$\left| t_n \right|$ (the modal significance) for the cases of two dipoles with the same length $\ell=\ell_1$ and dissimilar lengths $\ell_1=0.99\ell$ are depicted in the top and bottom panels of Fig.~\ref{fig:DipolesAbsT_final}, respectively. The corresponding phase angles~$\alpha_n$ (phase of eigenvalue~$t_n$) are shown in Fig.~\ref{fig:DipolesArgS}. Two characteristic modes (even and odd modes) dominate around the first resonance $\ell\approx 0.5\lambda$ as illustrated by black and red solid curves. Higher-order modes have very low modal significance~$\left| t_n \right| \approx 0$ and ill-defined phase~$\alpha_n$ and are not shown in the figures. The frequency interval~$l/\lambda \in [0.44,0.52]$ is sampled with 200 sample points and the result of tracking~\eqref{eq:track3} shows that eigentraces cross (in phase and in magnitude) for $\ell_1=\ell$ but do not cross (in phase) for $\ell_1=0.99\ell_1$ in agreement with the von Neumann-Wigner theorem~\cite{vonNeumannWigner_OnTheBehaviourOfEigenvaluesENG,SchabBernhard_GroupTheoryForCMA,Maseketal_ModalTrackingBasedOnGroupTheory}. The differences between eigenvalues and eigenvectors for the two cases are negligible away from the resonance at $\ell/ \lambda\approx 0.48$ and the tracking algorithm requires sample points close to the resonance to distinguish the two cases. 

Dipoles with non-zero surface resistivity are used to illustrate properties of characteristic modes for lossy structures. Lossless and low-loss cases are indistinguishable in the phase $\alpha_n$, as seen in Fig.~\ref{fig:DipolesArgS}, but a small amplitude decrease is visible in modal significance $|t_n|$. The differences increase with higher losses and the case with $R_{\T{s}}=0.1\unit{\Omega/\square}$ is clearly noticeable. Here, the eigentraces (both in phase and magnitude) of equal-lengths and different-lengths cases are similar  (dashed curves), both presenting crossing-avoidances dominated by very different values of modal significance of the even and odd mode at the  resonance frequency. By means of~\eqref{eq:Plost1}, this can be attributed to different modal radiation efficiencies~$\eta_n = \left(1 + \delta_n \right)^{-1} = - \left| t_n \right| / \cos \left( \alpha_n \right)$ of the even and odd modes, since at the external resonance~$\cos \left( \alpha_n \right) = - 1$ the modal radiation efficiency equals the modal significance.

\begin{figure}[]
    \centering
    \includegraphics[width=1\columnwidth]{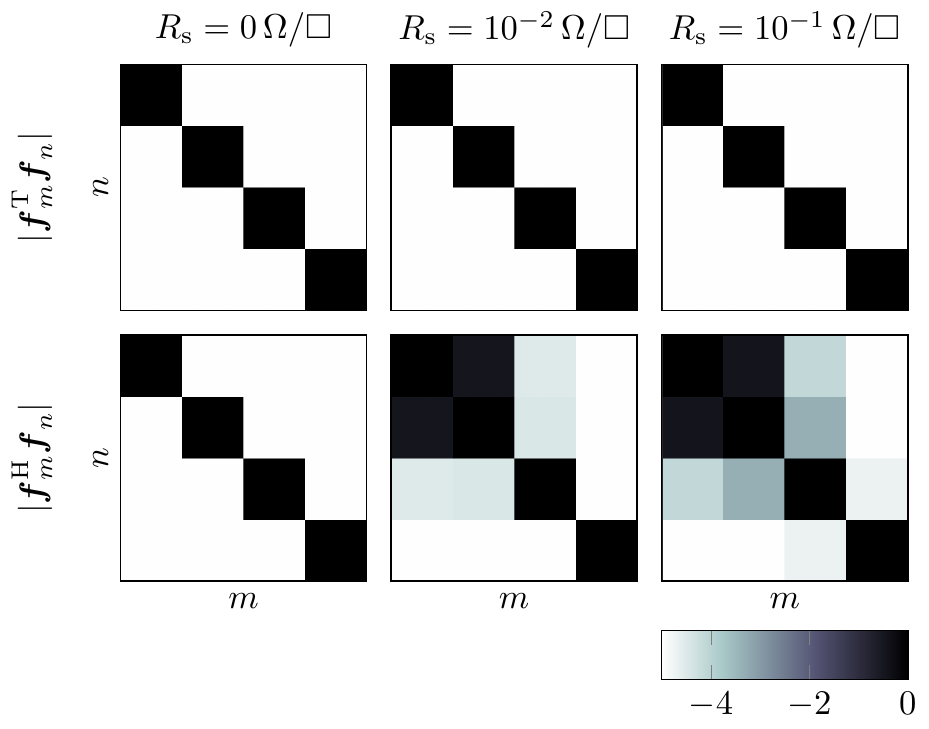}
    \caption{The reaction products $|\M{f}_m^\T{T}\M{f}_n|$ (top) and correlation coefficients $|\rho_{mn}| = |\M{f}_m^\herm\M{f}_n |$ (bottom) for the first four characteristic modes at resonant frequency of the dipoles with lengths $\ell_1=0.99\ell$.}
    \label{fig:DipoleCorr0}
\end{figure}

\begin{figure}[]
    \centering
    \includegraphics[width=\columnwidth]{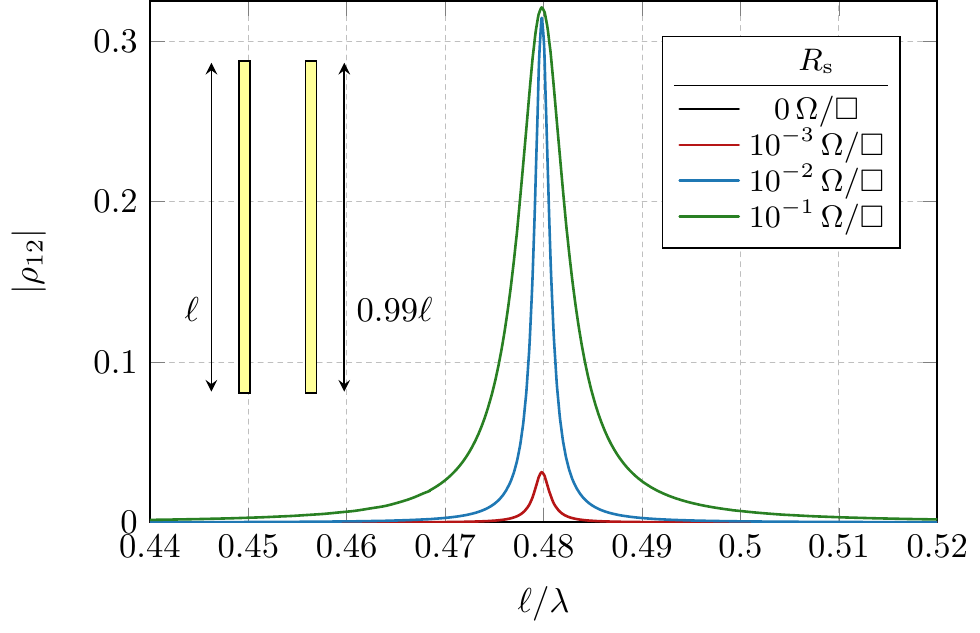}
    \caption{Correlation coefficient between the two dominant characteristic modes of two dipoles with length $\ell$ and $0.99\ell$ and different values of surface resistivity.}
    \label{fig:DipoleCorr}
\end{figure}

Losses also affect orthogonality of the far fields, here represented by the eigenvectors $\M{f}_n$. In reciprocal scenarios, the modal eigenvectors always satisfy relation~\eqref{eq:ff_reaction}, \ie{}, modes belonging to different eigenvalues exhibit vanishing reaction product, see the top row of in Fig.~\ref{fig:DipoleCorr0}. In the general lossy scenario, the modes are however not orthogonal as can be seen from the bottom row of Fig.~\ref{fig:DipoleCorr0} which depicts correlation coefficients~$\rho_{mn}$ given by~\eqref{eq:rho}.  The reaction product is diagonal for all considered cases. The matrix of correlation coefficients, however, is diagonal for the lossless case whereas the lossy cases have non-negligible correlation between the two dominant modes. This correlation $\varrho_{12}$ is detailed in Fig.~\ref{fig:DipoleCorr} around the resonance frequency, where it is seen that the correlation is $|\rho_{12}|\approx 0.3$ at $\ell=0.48\lambda$ for $R_{\T{s}}\in\{0.01,0.1\}\unit{\Omega/\square}$ but vanishes quickly away from the resonance. The correlation matrix for the $\ell_1=\ell$ case is approximately diagonal around the resonance but has non-diagonal elements for higher order modes at higher frequencies. In that case, the orthogonality, \ie{}, $\rho_{12} = 0$, follows from the symmetry of the structure which divides the dominant modes into different irreducible representations~\cite{McWeeny_GroupTheory}.       

\subsection{Discussion}
\label{sec:disc:lossy}

There is no consensus in the literature on how to define characteristic modes for lossy objects, and it is not possible to preserve all the convenient properties from the lossless case. As shown in Section~\ref{sec:CML} diagonalization of matrix~$\M{T}$ (or similarly matrix~$\M{Z}$) is in general not possible with modes having orthogonal far fields. However, numerical results in Section~\ref{sec:resDipole} indicate that ohmic losses commonly used to model metallic antennas only slightly perturb the modes and produce modes with low correlation~\eqref{eq:rho}, see also~\cite{YlaOijala_GeneralizedTCM2019,YlaOijala+Wallen2021}. Moreover, the characteristic mode superposition detailed in Section~VII-C of Part~I~\cite{Gustafsson+etal_CMT1_2021} only requires diagonal reaction products~\eqref{eq:ff_reaction}, which holds for lossy materials~\cite{HarringtonMautz_TheoryOfCharacteristicModesForConductingBodies}. 
 
Alternative possibilities to diagonalize matrices include the Autonne–Takagi factorization, which for reciprocal~$\M{T}$ is~$\M{A}^{\trans}\M{T}\M{A} = \M{\Lambda}$ with a unitary matrix~$\M{A}$ and a diagonal matrix~$\M{\Lambda}$ having the singular values of $\M{T}$ on the diagonal. This constructs modes with orthogonal far fields but without many of the good properties from ordinary diagonalization~\cite{HornJohnson_MatrixAnalysis}, such that summation formulas detailed in Section~VII-C of Part~I~\cite{Gustafsson+etal_CMT1_2021}. Autonne–Takagi factorization is similar to diagonalization of~$\M{T}^{\herm}\M{T}$ as proposed in~\cite{InagakiGarbacz_EigenfunctionsOfCompositeHermitianOperatorsWithApplicationToDiscreteAndContunuousRadiatingSystems} and discussed for lossy cases in~\cite{SarkarMokoleSalazarPalma_AnExposeOnInternalResonancesCM}. Similar properties are also obtained from diagonalization of~$\T{Re} \left\{ \M{T} \right\}$.

Generalizations directly based on the impedance matrix have also been investigated. In~\cite{HarringtonMautz_TheoryOfCharacteristicModesForConductingBodies}, it is shown that using the resistive part of the impedance matrix as a weighting matrix lead to modes with correlated far fields and modes with a combination of high losses and radiation. Radiation modes~\cite{Schab_PhDThesis,2020_Gustafsson_NJP} is an alternative which produces modes with orthogonal far fields but independent of the reactance.

\section{Characteristic modes based on FEM}
\label{S:FEM}
Characteristic modes based on the transition matrix~\eqref{eq:CM5} are independent of the underlying numerical algorithm used to compute the matrix~$\M{T}$. Method of moments used in the previous examples and the theory developed in part~I~\cite{Gustafsson+etal_CMT1_2021} dominates literature on characteristic modes~\cite{HarringtonMautz_ComputationOfCharacteristicModesForConductingBodies,HarringtonMautzChang_CharacteristicModesForDielectricAndMagneticBodies,ChenWang_CharacteristicModesWiley} to the extent that it is sometimes claimed that characteristic mode decomposition is only possible in MoM-based formulations~\cite{SarkarMokoleSalazarPalma_AnExposeOnInternalResonancesCM}. That this is not the case is demonstrated in this section showing a formulation of characteristic modes based on the theory developed in Part~I~\cite{Gustafsson+etal_CMT1_2021} with a transmission matrix~$\M{T}$ computed using the finite element method (FEM)~\cite{Demesy+etal2018,Fruhnert+etal2017}. This formulation also sheds light on properties of the incident field encoded in the projection matrix~$\M{U}$ which shows rapid decay of electric field energy inside a penetrable scatterer for higher-order spherical modes, ordered by degree of spherical vector waves. This motivates the estimate of the maximum degree given in~\eqref{eq:Lmax}. 

\begin{figure}%
\centering
\begin{tikzpicture}
\node[inner sep=0pt,anchor=south west] at (0,0)
    {\includegraphics[width=36mm]{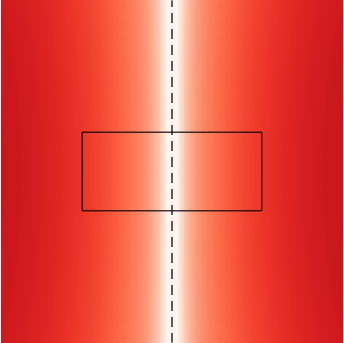}};
\pgfplotscolorbardrawstandalone[ 
		colormap name={OrRd-9},
    colorbar horizontal,
    point meta min=0,
    point meta max=1,
    colorbar style={
				at={(18mm,41mm)},
        height=3mm,
        width=20mm,
				anchor=south,
        xtick={0,1}, 
				xticklabels = {$-50$\,dB,$-10$\,dB},
				x tick label style={yshift=2ex,anchor=south}}]
\node[right] at (2mm,33mm) {TE};				
\node[right] at (2mm,29mm) {$l=1$};				
\end{tikzpicture}
\hspace{1mm}
\begin{tikzpicture}
\node[inner sep=0pt,anchor=south west] at (0,0)
    {\includegraphics[width=36mm]{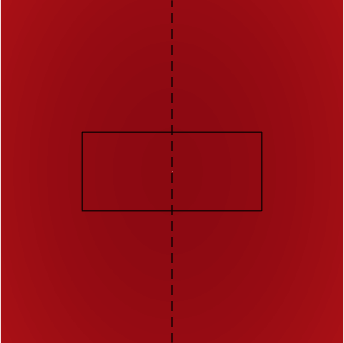}};
\pgfplotscolorbardrawstandalone[ 
		colormap name={OrRd-9},
    colorbar horizontal,
    point meta min=0,
    point meta max=1,
    colorbar style={
				at={(18mm,41mm)},
        height=3mm,
        width=20mm,
				anchor=south,
        xtick={0,1}, 
				xticklabels = {$-50$\,dB,$-10$\,dB},
				x tick label style={yshift=2ex,anchor=south}}]
\node[right] at (2mm,33mm) {TM};				
\node[right] at (2mm,29mm) {$l=1$};								
\end{tikzpicture}
\vspace{2mm}

\begin{tikzpicture}
\node[inner sep=0pt,anchor=south west] at (0,0)
    {\includegraphics[width=36mm]{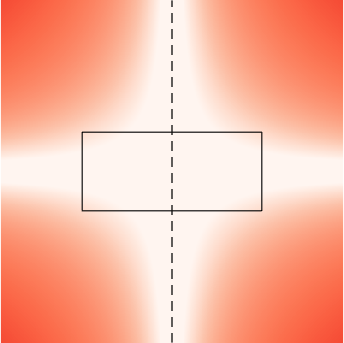}};
\pgfplotscolorbardrawstandalone[ 
		colormap name={OrRd-9},
    colorbar horizontal,
    point meta min=0,
    point meta max=1,
    colorbar style={
				at={(18mm,41mm)},
        height=3mm,
        width=20mm,
				anchor=south,
        xtick={0,1}, 
				xticklabels = {$-50$\,dB,$-10$\,dB},
				x tick label style={yshift=2ex,anchor=south}}]
\node[right] at (2mm,33mm) {TE};				
\node[right] at (2mm,29mm) {$l=2$};								
\end{tikzpicture}
\hspace{1mm}
\begin{tikzpicture}
\node[inner sep=0pt,anchor=south west] at (0,0)
    {\includegraphics[width=36mm]{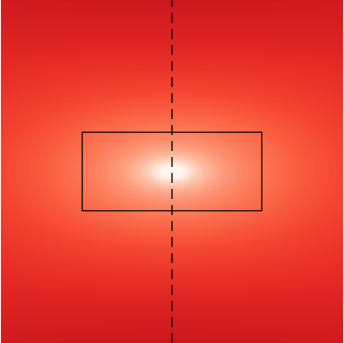}};
\pgfplotscolorbardrawstandalone[ 
		colormap name={OrRd-9},
    colorbar horizontal,
    point meta min=0,
    point meta max=1,
    colorbar style={
				at={(18mm,41mm)},
        height=3mm,
        width=20mm,
				anchor=south,
        xtick={0,1}, 
				xticklabels = {$-50$\,dB,$-10$\,dB},
				x tick label style={yshift=2ex,anchor=south}}]
\node[right] at (2mm,33mm) {TM};				
\node[right] at (2mm,29mm) {$l=2$};								
\end{tikzpicture}

\caption{Amplitudes of incident electric fields used to compute the transmission elements $\M{T}$ for azimuth index $m=0$ (field independent of $\varphi$) in the $y=0$ plane and $|x|\leq 10\unit{mm}$ and $|z|\leq 10\unit{mm}$ for frequency~$f = 4.87\unit{GHz}$. The cross-section of the dielectric cylinder is illustrated with the black rectangle and the axis of symmetry with the dashed line. Spherical wave indices~$l=1$ (top row), $l=2$ (bottom row), TE (left), and TM (right) are shown.}%
\label{fig:CylEinSphWaves}%
\end{figure}
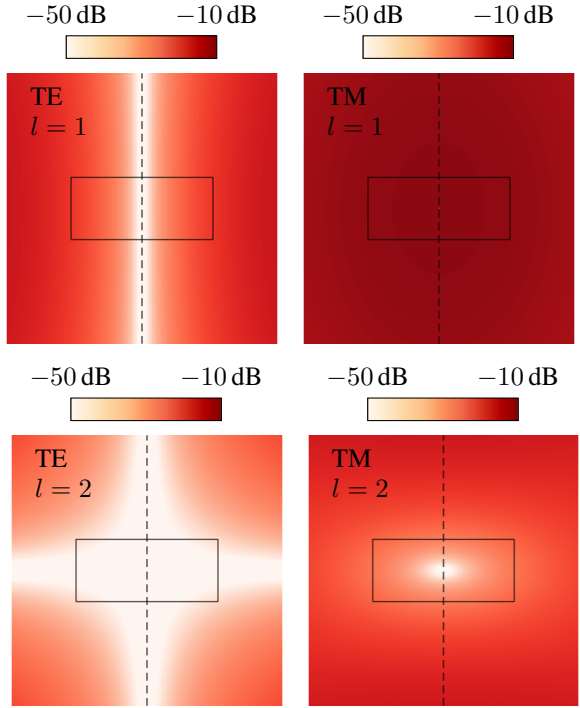 

Consider the dielectric cylinder used in Section~\ref{S:ResCyl} around the resonance frequency~$f = 4.87\unit{GHz}$ of the $\T{TE}_{01}$ mode, see Fig.~\ref{fig:Cyl38absT}. This mode corresponds to a magnetic dipole $l=1$ with a $\hat{\V{\varphi}}$-directed electric field which is independent of the azimuth angle $\varphi$ ($m=0$, index $m$ being the analogue of magnetic quantum number attached to spherical harmonics~\cite{Kristensson_ScatteringBook}). The transition matrix~$\M{T}$ is calculated by illuminating the dielectric cylinder with electromagnetic fields corresponding to isolated spherical vector waves and subsequent decomposition of the resulting total field into the same basis using the projection matrix ~$\M{U}$, see part~I~\cite{Gustafsson+etal_CMT1_2021}. Four spherical waves depicted in Fig.~\ref{fig:CylEinSphWaves} were used in this case, resulting in a matrix~$\M{T}$ of size~$4 \cross 4$. These incident fields are depicted over a $40\unit{dB}$ dynamic range, and the TM dipole mode ($l=1$) has the highest amplitude in the region of the cylinder. Electric field of TE modes vanishes at the symmetry axis ($x=0$ and $y=0$) and the amplitude of electric field decreases in the region of the cylinder with increasing modal index $l$, in agreement with the rule of thumb that modes with order smaller than $\lceil k\aCircum\rceil$ dominate~\cite{Hansen_SphericalNearFieldAntennaMeasurements} the spherical expansion, where we note that electrical size~$k\aCircum\approx 0.56$ at frequency~$f = 4.87\unit{GHz}$. 

\begin{figure}%
\centering
\begin{tikzpicture}
\node[inner sep=0pt,anchor=south west] at (0,0)
    {\includegraphics[width=36mm]{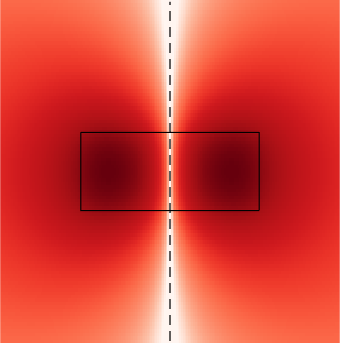}};
\pgfplotscolorbardrawstandalone[ 
		colormap name={OrRd-9},
    colorbar horizontal,
    point meta min=0,
    point meta max=1,
    colorbar style={
				at={(18mm,41mm)},
        height=3mm,
        width=20mm,
				anchor=south,
        xtick={0,1}, 
				xticklabels = {$-30$\,dB,$10$\,dB},
				x tick label style={yshift=2ex,anchor=south}}]
\node[right] at (2mm,33mm) {TE};				
\node[right] at (2mm,29mm) {$l=1$};				
\end{tikzpicture}
\hspace{1mm}
\begin{tikzpicture}
\node[inner sep=0pt,anchor=south west] at (0,0)
    {\includegraphics[width=36mm]{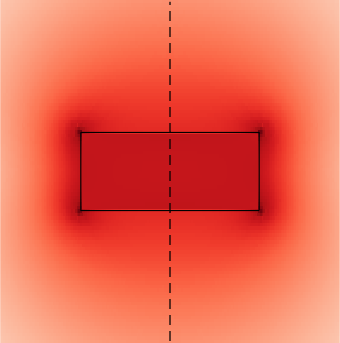}};
\pgfplotscolorbardrawstandalone[ 
		colormap name={OrRd-9},
    colorbar horizontal,
    point meta min=0,
    point meta max=1,
    colorbar style={
				at={(18mm,41mm)},
        height=3mm,
        width=20mm,
				anchor=south,
        xtick={0,1}, 
				xticklabels = {$-40$\,dB,$0$\,dB},
				x tick label style={yshift=2ex,anchor=south}}]
\node[right] at (2mm,33mm) {TM};				
\node[right] at (2mm,29mm) {$l=1$};								
\end{tikzpicture}
\vspace{1mm}

\begin{tikzpicture}
\node[inner sep=0pt,anchor=south west] at (0,0)
    {\includegraphics[width=36mm]{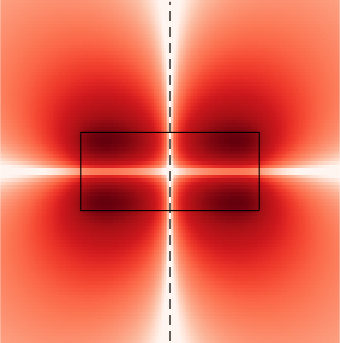}};
\pgfplotscolorbardrawstandalone[ 
		colormap name={OrRd-9},
    colorbar horizontal,
    point meta min=0,
    point meta max=1,
    colorbar style={
				at={(18mm,41mm)},
        height=3mm,
        width=20mm,
				anchor=south,
        xtick={0,1}, 
				xticklabels = {$-90$\,dB,$-50$\,dB},
				x tick label style={yshift=2ex,anchor=south}}]
\node[right] at (2mm,33mm) {TE};				
\node[right] at (2mm,29mm) {$l=2$};								
\end{tikzpicture}
\hspace{2mm}
\begin{tikzpicture}
\node[inner sep=0pt,anchor=south west] at (0,0)
    {\includegraphics[width=36mm]{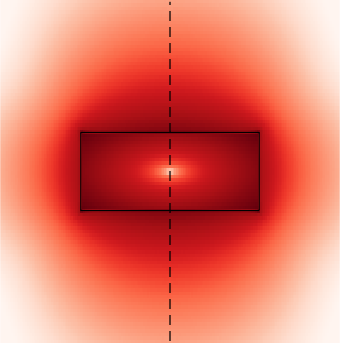}};
\pgfplotscolorbardrawstandalone[ 
		colormap name={OrRd-9},
    colorbar horizontal,
    point meta min=0,
    point meta max=1,
    colorbar style={
				at={(18mm,41mm)},
        height=3mm,
        width=20mm,
				anchor=south,
        xtick={0,1}, 
				xticklabels = {$-65$\,dB,$-25$\,dB},
				x tick label style={yshift=2ex,anchor=south}}]
\node[right] at (2mm,33mm) {TM};				
\node[right] at (2mm,29mm) {$l=2$};								
\end{tikzpicture}
\caption{Amplitudes of the scattered electric field (independent of $\varphi$) for the dielectric cylinder from the incident fields in Fig.~\ref{fig:CylEinSphWaves} evaluated in the $y=0$ plane for $|x|\leq 10\unit{mm}$ and $|z|\leq 10\unit{mm}$. $l=1$ and $l=2$ in top and bottoms rows.}%
\label{fig:CylEscattSphWaves}%
\end{figure}
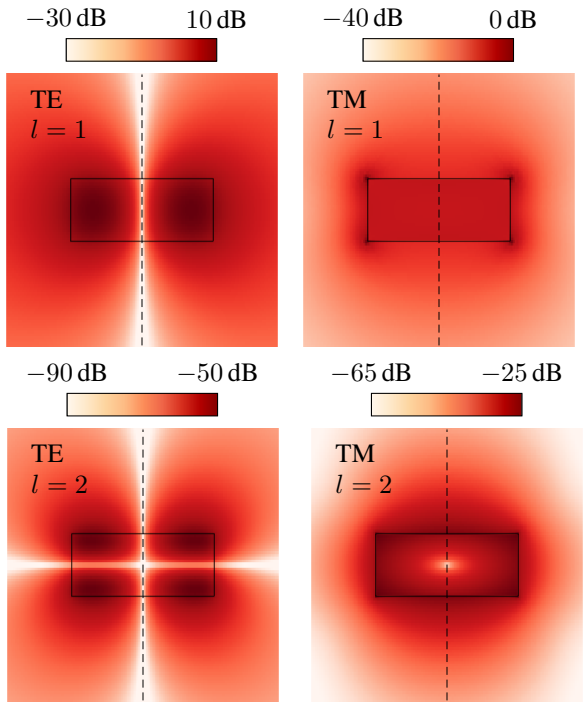 

Scattered fields are computed using FEM\footnote{COMSOL Multiphysics~5.6 with axial symmetry and a spherical perfectly matched layer (PML) with inner radius $60~\T{mm}$ was used.} with the incident fields from Fig.~\ref{fig:CylEinSphWaves}. The computed scattered electric fields are depicted in Fig.~\ref{fig:CylEscattSphWaves} for the $y=0$ plane restricted to $|x|\leq 10\unit{mm}$ and $|z|\leq 10\unit{mm}$. The scattered field shows a dominant transfer of incident $\T{TE}_{01}$ mode into scattered $\T{TE}_{01}$ mode. Other modal combinations are negligible. The $\T{TM}_{01}$ mode has a large incident field, but this mode gives important transfer into scattered field only at resonance frequency~$f = 7.53\unit{GHz}$ as seen in Fig.~\ref{fig:Cyl38absT}. Hence for this example, there exists only one dominant eigenmode in the solution of~\eqref{eq:CM5}.

The results here represent only the $m=0$ block of the matrix $\M{T}$.  The complete matrix~$\M{T}$ also contains elements corresponding to $\phi$-dependent fields ($|m|>0$) computed analogously, but for ease of presentation those data are not depicted here. By the block nature of the matrix $\M{T}$ for this axially symmetric problem, eigenvalue decomposition of $|m|>0$ blocks will only affect eigenvalues and eigenvectors of higher order modes with corresponding axial variation.

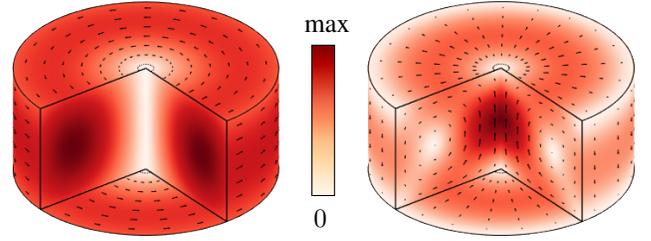
\begin{figure}%
\centering
\hspace{1mm}
\includegraphics[width=0.4\columnwidth]{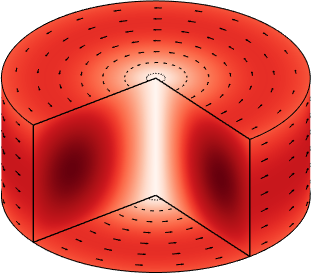}%
\hspace{1mm}
\begin{tikzpicture}
\pgfplotscolorbardrawstandalone[ 
		colormap name={OrRd-9},
    colorbar,
    point meta min=0,
    point meta max=1,
    colorbar style={
				at={(0,0)},
        height=20mm,
        width=3mm,
				anchor=south,
        ytick={2}}
				]
\node[right] at (-0.75mm,22.5mm) {max};				
\node[right] at (0.5mm,-3mm) {$0$};
\end{tikzpicture}
\includegraphics[width=0.4\columnwidth]{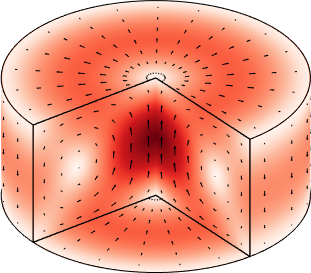}%
\caption{Characteristic currents for the $\T{TE}_{01}$ mode (left) at frequency~$f = 4.87\unit{GHz}$ and the $\T{TM}_{01}$ mode (right) at frequency~$f = 7.53\unit{GHz}$ in the dielectric cylinder calculated using FEM~\eqref{eq:FEMJ}, \textit{cf.} the corresponding equivalent surface current densities in Fig.~\ref{fig:Cyl38JM}. Absolute value of current density is plotted in linear scale normalized with the maximum magnitude.}%
\label{fig:Cyl38JFEM}%
\end{figure} 

Characteristic currents~$\V{J}_n$ corresponding to an eigenvector~$\M{f}_n$ are computed in FEM from the total electric field~$\V{E}_n$ induced by an incident field given by the spherical wave composition encoded in eigenvector~$\M{a}_n=t_n^{-1}\M{f}_n$ as the contrast (polarization) current
\begin{equation}
	\V{J}_n(\V{r}) = \T{j}kZ^{-1}\big(\varepsilon_{\T{r}}(\V{r})-1\big)\V{E}_n(\V{r}),
\label{eq:FEMJ}
\end{equation}
where~$k$ is the wavenumber, $Z$ wave impedance, and~$\varepsilon_\T{r}$ the relative permittivity. The characteristic currents for the resonant modes ($\T{TE}_{01}$) at frequency~$f = 4.87\unit{GHz}$ and ($\T{TM}_{01}$) at frequency~$f = 7.53\unit{GHz}$ are depicted in Fig.~\ref{fig:Cyl38JFEM}. This current density corresponds to the equivalent current densities shown in Fig.~\ref{fig:Cyl38JM}, see also~\cite{2016_Hu_TAP,Ylae-Oijala2018,Huang+etal2019}.

\subsection{Discussion}
Characteristic modes based on the transmission matrix enable computation tools such as FEM to complement classical \ac{MoM}-based formulations. In contrast to \ac{MoM}, where the current is discretized, fields supported in a volume surrounding the object are discretized, and absorbing boundary conditions are used to terminate the mesh in FEM~\cite{Jin_TheoryAndComputationOfElectromagneticFields}. These differences suggest that the use of FEM is detrimental in the computation of characteristic modes, but further studies are needed to investigate the accuracy, computational cost, and practical use of FEM-based CM.  

\section{Conclusion}
\label{sec:concl}

The unified theory of characteristic modes developed in Part~I was applied in this Part~II to accelerate modal tracking and make it more precise, to address the lack of far-field orthogonality for lossy characteristic modes, and to demonstrate that the knowledge of transition matrix makes it possible to evaluate characteristic modes irrespective of what numerical method is used, rendering the scattering-based approach a universal and standalone frequency-domain technique.

The interpolation and fitting methods enabled by the properties of the transition matrix allows for a sizeable reduction in the number of single-frequency calculations required to synthesize accurate broadband characteristic mode data. Performed over eigenvectors of transition matrix, the tracking is inherently done for modal far fields, without necessity to calculate them during postprocessing step from modal currents. Tracking is facilitated here by employing Foster's reactance theorem and properties of scattering matrix.

The possibility of defining characteristic modes for lossy obstacles, a popular topic in the recent years, was discussed in detail, including a thorough exposition and numerical evidence why it is not possible to diagonalize impedance matrix and preserve far-field orthogonality in a lossy scenario. It has been found that this is a fundamentals issue which holds for all existing formulations.

Finally, the finite element method was employed to compute characteristic fields of a dielectric obstacle demonstrating that the scattering-based formulation of characteristic modes is compatible with any numerical method capable of producing transition matrix data.

\appendices

\section{Interpolation between unitary matrices}
\label{S:IntUnitaryMat}
The interpolation~\eqref{eq:track1} 
\begin{equation}
    \M{F} = \M{A}(\M{A}^{-1}\M{B})^\nu
    =\M{A}\M{C}^\nu, \, \nu \in [0, 1]
\end{equation}
is an interpolation between two unitary matrices $\M{A}$ and $\M{B}$.  Both matrices~$\M{C}=\M{A}^{-1}\M{B}$ and~$\M{F}$ are also unitary, apparent from
\begin{equation}
\begin{aligned}
    \M{C}^{\herm}\M{C}&=\M{B}^{\herm}\M{A}^{-\herm}\M{A}^{-1}\M{B} = \M{1} \\ 
    \M{F}^{\herm}\M{F}
	&=(\M{C}^{\nu})^{\herm}\M{A}^{\herm}\M{A}\M{C}^{\nu}
	=(\M{C}^{\herm})^{\nu}\M{C}^{\nu}
	=(\M{C}^{\herm}\M{C})^{\nu}
	=\M{1}.
\end{aligned}    
\end{equation}
In all cases, the matrix exponent is computed from a diagonalization 
\begin{equation}
    \M{C}^{\nu}=(\M{U}\M{D}\M{U}^{\herm})^\nu
    =\M{U}\M{D}^{\nu}\M{U}^{\herm}
\end{equation}
with a diagonal matrix~$\M{D}$ and a unitary matrix~$\M{U}$. The interpolation can also be expressed using the matrix exponent and matrix logarithm
\begin{equation}
    \M{C}^{\nu}=\exp(\nu\log(\M{C}))
\end{equation}
which highlights the connection with linear interpolation of the phase. This is clearly seen in the scalar case with $A=\exp(\J\phi_{\T{a}})$ and $B=\exp(\J\phi_{\T{b}})$ giving
\begin{equation}
	F = A^{1-\nu}B^\nu
	=A(B/A)^\nu
	=\eu^{\J (\phi_{\T{a}}(1-\nu)+\nu\phi_{\T{b}})}.
\end{equation}

\section{Simultaneous diagonalization of three matrices}
\label{S:DiagTreeMat}
Diagonalization by Hermitian congruence~\cite{HornJohnson_MatrixAnalysis}
\begin{equation}
    \M{X}^{\herm}(\M{A}+\J\M{B})\M{X}=\M{D}
    \quad\text{and }
    \M{X}^{\herm}\M{C}\M{X}=\M{1}
\end{equation}
of three real symmetric matrices $\M{A}$, $\M{B}$, and $\M{C}$ is equivalent to simultaneous diagonalization of the three matrices. Here, $\M{D}$ denotes a diagonal matrix and, without loss of generality, $\M{X}$ is normalized according to the second diagonalization. Assume that $\M{C}=\M{U}^{\trans}\M{U}=\M{U}^{\herm}\M{U}$ is positive definite and substitute the unitary matrix $\M{Y}=\M{U}\M{X}$ into the first diagonalization
\begin{equation}
    \M{Y}^{\herm}\big(\M{U}^{-\herm}(\M{A}+\J\M{B})\M{U}^{-1}\big)\M{Y}=\M{D}.
\end{equation}
Using that diagonalization with a unitary matrix is equivalent with normality~\cite{HornJohnson_MatrixAnalysis} shows that  
\begin{equation}
    \M{U}^{-\herm}(\M{A}+\J\M{B})\M{U}^{-1}
\end{equation}
is normal matrix which reduces to a necessary and sufficient condition for simultaneous diagonalization of three matrices~\cite{Novikov2014}
\begin{equation}
	\M{A}\M{C}^{-1}\M{B} = \M{B}\M{C}^{-1}\M{A}.
\label{eq:congu}
\end{equation}
The involved matrices can be permuted as long as they are invertible, see~\cite{Novikov2014} for additional constraints for indefinite and singular matrices.  

\bibliographystyle{IEEEtran}
\bibliography{references,extraBib}


\begin{IEEEbiography}[{\includegraphics[width=1in,height=1.25in,clip,keepaspectratio]{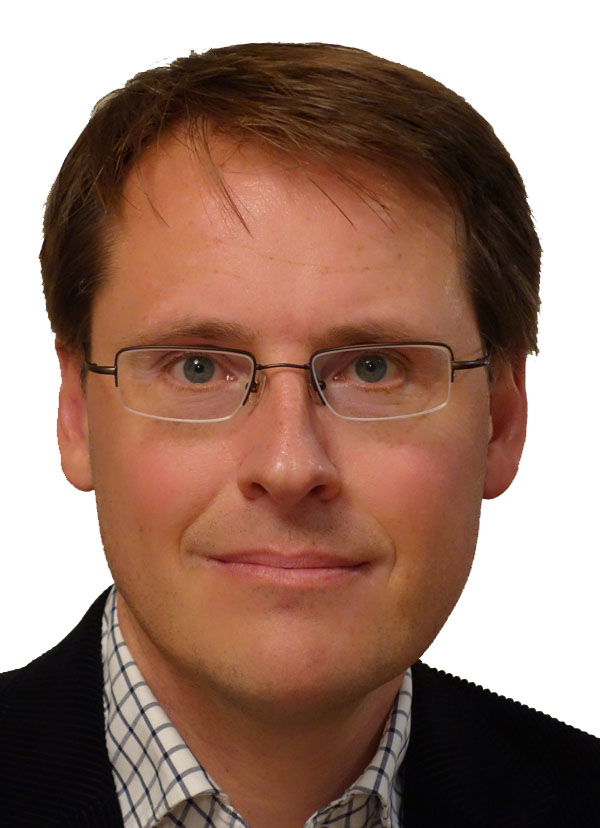}}]{Mats Gustafsson}
received the M.Sc. degree in Engineering Physics 1994, the Ph.D. degree in Electromagnetic Theory 2000, was appointed Docent 2005, and Professor of Electromagnetic Theory 2011, all from Lund University, Sweden.

He co-founded the company Phase holographic imaging AB in 2004. His research interests are in scattering and antenna theory and inverse scattering and imaging. He has written over 100 peer reviewed journal papers and over 100 conference papers. Prof. Gustafsson received the IEEE Schelkunoff Transactions Prize Paper Award 2010, the IEEE Uslenghi Letters Prize Paper Award 2019, and best paper awards at EuCAP 2007 and 2013. He served as an IEEE AP-S Distinguished Lecturer for 2013-15.
\end{IEEEbiography}

\begin{IEEEbiography}[{\includegraphics[width=1in,height=1.25in,clip,keepaspectratio]{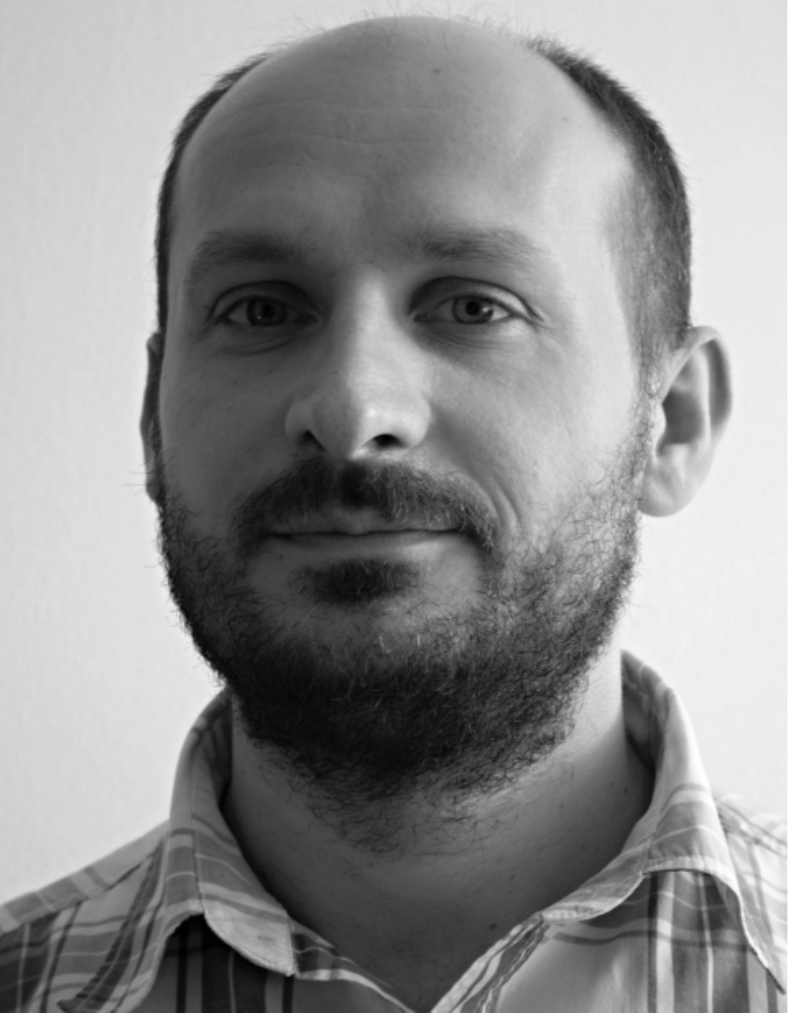}}]{Lukas Jelinek}
received his Ph.D. degree from the Czech Technical University in Prague, Czech Republic, in 2006. In 2015 he was appointed Associate Professor at the Department of Electromagnetic Field at the same university.

His research interests include wave propagation in complex media, electromagnetic field theory, metamaterials, numerical techniques, and optimization.
\end{IEEEbiography}

\begin{IEEEbiography}[{\includegraphics[width=1in,height=1.25in,clip,keepaspectratio]{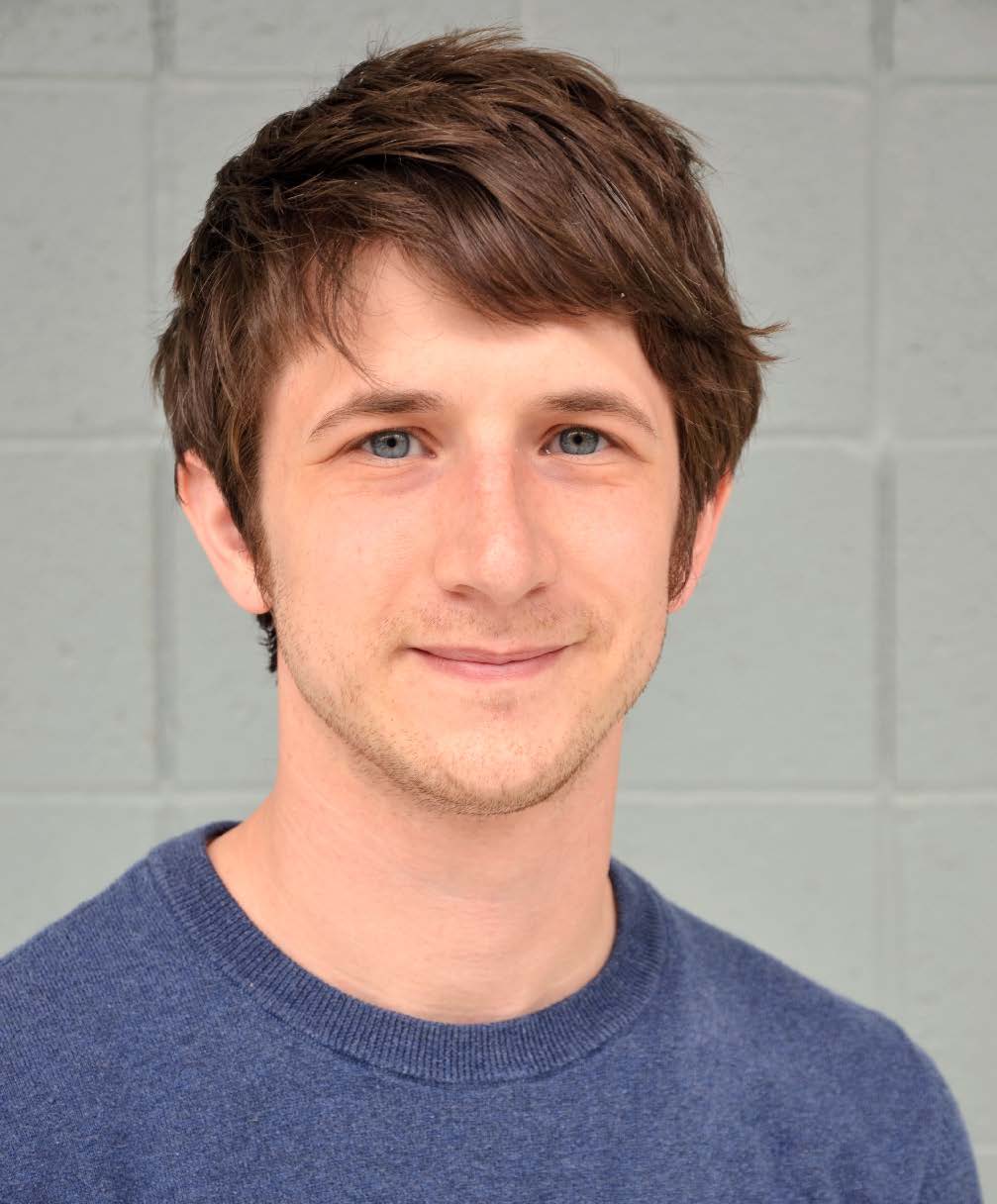}}]{Kurt Schab}(M'16)
Kurt Schab is an Assistant Professor of Electrical Engineering at Santa Clara University, Santa Clara, CA USA. He received the B.S. degree in electrical engineering and physics from Portland State University in 2011, and the M.S. and Ph.D. degrees in electrical engineering from the University of Illinois at Urbana-Champaign in 2013 and 2016, respectively.  From 2016 to 2018 he was an Intelligence Community Postdoctoral Research Scholar at North Carolina State University in Raleigh, North Carolina.  His research focuses on the intersection of numerical methods, electromagnetic theory, and antenna design.  
\end{IEEEbiography}

\begin{IEEEbiography}[{\includegraphics[width=1in,height=1.25in,clip,keepaspectratio]{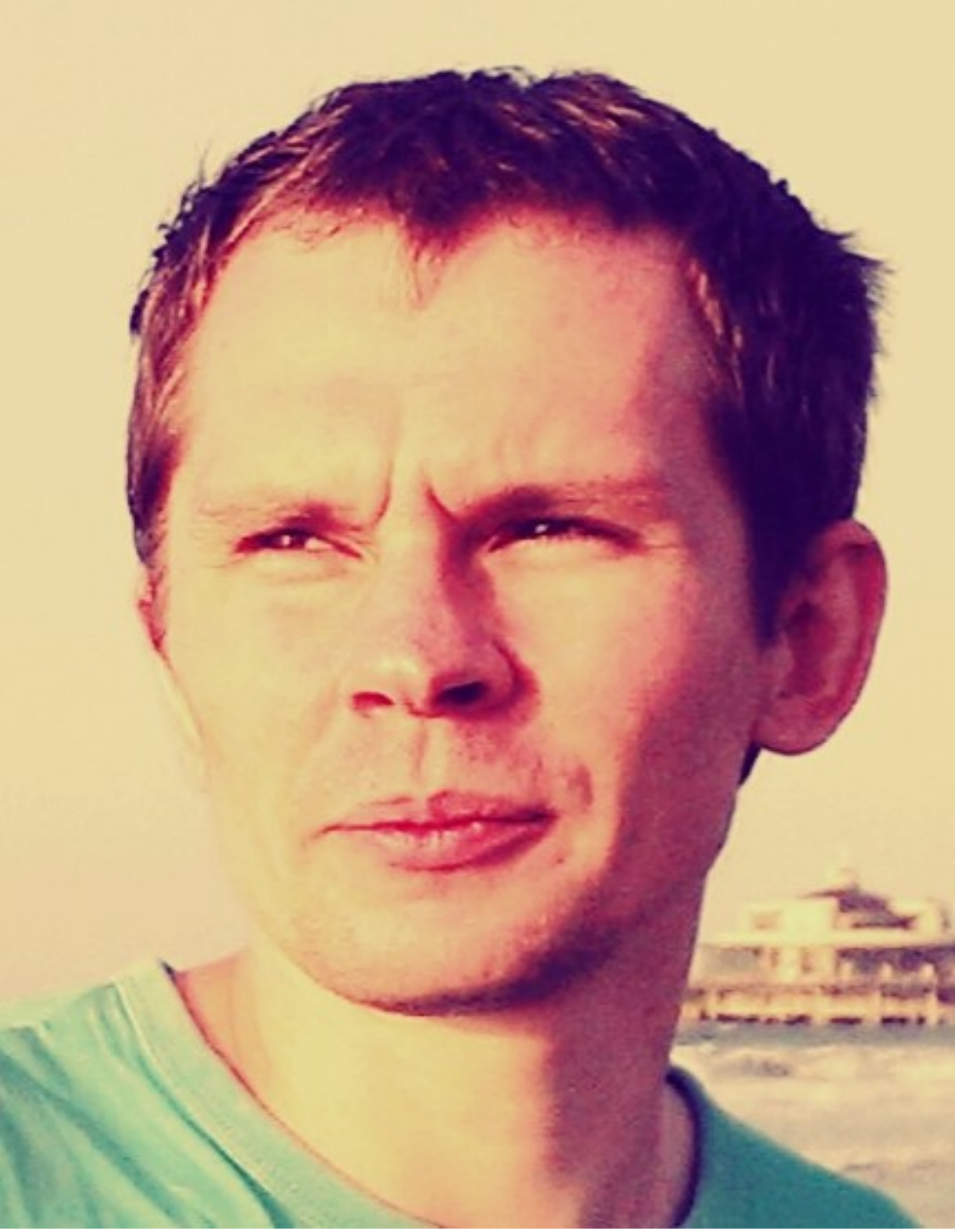}}]{Miloslav Capek}
(M'14, SM'17) received the M.Sc. degree in Electrical Engineering 2009, the Ph.D. degree in 2014, and was appointed Associate Professor in 2017, all from the Czech Technical University in Prague, Czech Republic.
	
He leads the development of the AToM (Antenna Toolbox for Matlab) package. His research interests are in the area of electromagnetic theory, electrically small antennas, numerical techniques, and optimization. He authored or co-authored over 100~journal and conference papers.

Dr. Capek is member of Radioengineering Society and Associate Editor of IET Microwaves, Antennas \& Propagation.
\end{IEEEbiography}

\end{document}